\documentclass[preprint,12pt]{elsarticle}

\usepackage{amssymb}
\usepackage{graphicx}
\usepackage{amssymb,amsfonts,amsmath}
\usepackage[figuresright]{rotating}
\usepackage[normalem]{ulem}

\usepackage{subcaption}

\usepackage{booktabs}

\journal{arXiv}

\usepackage{color}
	 \definecolor{darkred}{rgb}{0.75,0,0}
	 \definecolor{darkgreen}{rgb}{0,0.5,0}
	 \definecolor{darkblue}{rgb}{0,0,0.75}

\usepackage{hyperref}
\definecolor{darkgreen}{rgb}{0.1,0.6,0.3}
\definecolor{darkred}{rgb}{0.6,0.3,0.1}
\hypersetup{
colorlinks=true,   
linkcolor=blue,  
citecolor=darkgreen,
filecolor=magenta,  
urlcolor= black   
}

\usepackage[T1]{fontenc}

\begin{document}


\begin{frontmatter}



\title{Compartment model of strategy-dependent time delays in replicator dynamics}

\author[inst1]{Ma\l{}gorzata Fic\corref{cor1}\fnref{inst2}}
\ead{fic@evolbio.mpg.de}
 \cortext[cor1]{Corresponding author:}
\affiliation[inst1]{organization={Department of Theoretical Biology, Max Planck Institute for Evolutionary Biology},
addressline={August-Thienemann-Str. 2},
city={Ploen},
postcode={24306},
country={Germany}}
\affiliation[inst2]{organization={Center for Computational and Theoretical Biology, University of Wuerzburg},
addressline={ Klara-Oppenheimer-Weg 32},
city={Wuerzburg},
postcode={97074},
country={Germany}}

\author[inst3]{Frank Bastian}

\affiliation[inst3]{organization={School of Mathematical Sciences, University College Cork},
addressline={Western Road},
city={Cork},
postcode={T12 XF62},
country={Ireland}}

\author[inst4]{Jacek Mi\c{e}kisz}

\affiliation[inst4]{organization={Institute of Applied Mathematics and Mechanics, University of Warsaw},
addressline={ulica Banacha 2}, 
city={Warsaw},
postcode={02-097}, 
country={Poland}}

\author[inst2]{Chaitanya S. Gokhale\fnref{inst1}}

\begin{abstract}
\small{
Real-world processes often exhibit temporal separation between actions and reactions - a characteristic frequently ignored in many modelling frameworks. 
Adding temporal aspects, like time delays, introduces a higher complexity of problems and leads to models that are challenging to analyse and computationally expensive to solve. 
In this work, we propose an intermediate solution to resolve the issue in the framework of evolutionary game theory.
Our compartment-based model includes time delays while remaining relatively simple and straightforward to analyse. 
We show that this model yields qualitatively comparable results with models incorporating explicit delays. Particularly, we focus on the case of delays between parents' interaction and an offspring joining the population, with the magnitude of the delay depending on the parents' strategy.
We analyse Stag-Hunt, Snowdrift, and the Prisoner's 
Dilemma game and show that strategy-dependent delays are detrimental to affected strategies. 
Additionally, we present how including delays may change the effective games played in the population, subsequently emphasising the importance of considering the studied systems' temporal aspects to model them accurately.}
\end{abstract}




\end{frontmatter}



\section{Introduction}

	Evolutionary game theory provides a robust framework for modelling many biological and social interactions \cite{maynard-smith:Nature:1973}. 
	Traditional models largely exclude time delays, instead assuming that the actions of an individual are instantaneous and their impacts on fitness are immediate. 
	However, real-world processes frequently exhibit temporal separation between actions and their effects across scales of biological organisation. 
	Sporulation, the response to nutrient deprivation observed in bacteria like \emph{Bacillus subtilis}, takes about 8 to 10 hours to be completed after being triggered \cite{serra:JBA:2014}. 
	A seed bank in weeds leads to the sprouting of plants even years after the seed has been produced \cite{lauenroth:NP:2023}. 
	In the animal kingdom, the delays are also abundant. 
In the case of marine midges \emph{Clunio marinus}, the time between reproduction and emergence is correlated with the lunar cycle and takes from about 15 or 30 days, as compared to their relatively short adult lives of 2 hours \cite{kaiser:ME:2021}.
	Similarly, pregnancy and nurturing are temporally disconnected actions that culminate in adding a new, mature individual to the population \cite{bateson:TEE:1994}. 
	Social systems, too, display such delays, evident in long-term financial investments or latency in information transmission and processing \cite{binswanger:JEBO:2012, bourne:JEP:1957}. 
	Thus, incorporating temporal factors could refine evolutionary game theoretical models, capturing realistic dynamics more accurately.


The effects of time delays on replicator dynamics were
discussed in \cite{tao:JTB:1997, alboszta:JTB:2004, oaku:JER:2002, iijima:MSS:2011, iijima:JTB:2012, moreira:JTB:2012, miekisz:DGAA:2011, miekisz:DGA:2014, wesson:DGA:2016, wesson:IJBC:2016, khalifa:DGA:2018, khalifa:DGA:2016, bodnar:DGAA:2020, miekisz:PRE:2021, wettergren:AMPC:2023}. 
In \cite{tao:JTB:1997}, the authors introduced a social-type model in which individuals at time $t$ imitate a strategy with a higher average payoff at time $t-\tau$ for some time delay $\tau$. 
They showed that the interior stationary state of the resulting time-delayed differential equation is locally asymptotically stable for small time delays. 
In contrast, it becomes unstable for big ones, and oscillations appear.
In \cite{alboszta:JTB:2004}, a biological type model was constructed in which individuals are born $\tau$ units of time after their parents played.
This type of delay leads to a system of equations of strategy's frequency and population size.
The authors showed the absence of oscillations — the original stationary point is globally asymptotically stable for any time delay. 
Recently, models with strategy-dependent time delays were introduced \cite{moreira:JTB:2012, wesson:DGA:2016, khalifa:DGA:2016, bodnar:DGAA:2020, miekisz:PRE:2021}.

In \cite{bodnar:DGAA:2020, miekisz:PRE:2021}, the authors discussed the biological-type model of \cite{tao:JTB:1997} with strategy-dependent time delays. 
They reported a novel behaviour, showing that stationary states depend continuously on time delays. 
Moreover, at specific time delays, an interior stationary state may disappear, or another interior stationary state may appear. 
The equation for the stationary state of frequencies of strategies was derived and solved numerically in \cite{miekisz:PRE:2021} for Stag-Hunt, Snowdrift, and Prisoner's Dilemma games.
In \cite{miekisz:arXiv:2023}, a small-time delay approximation was proposed.
The authors derived an explicit formula for the stationary state, which approximates well the exact results of \cite{miekisz:PRE:2021}.


Here, we introduce a model that shares the analytical tractability of \cite{miekisz:arXiv:2023} without assuming small delays. 
We extend the approach of Mi\c{e}kisz and Bodnar \cite{miekisz:PRE:2021} and derive a system of ordinary differential equations that is easier to analyse through either closed-form solutions of stationary states or existing bifurcation software. 
In the model, time delays are represented by rates at which an offspring grows and can participate in games. 
More precisely, in our Kindergarten model, a newly born offspring is located in an inactive compartment. 
Then, with some strategy-dependent intensity, juveniles become players and move to an active compartment (this is reminiscent of models of delayed protein degradation in gene expression present in \cite{miekisz:BMB:2011}). 
This approach allows us to derive explicit analytical formulas for stationary states of the strategies. 
In Section~\ref{section:MM}, we recall the model of Mi\c{e}kisz and Bodnar \cite{miekisz:PRE:2021} and present our approach to model time delays. Results are described in Section~\ref{results}, and Conclusions follow in Section~\ref{conclusions}.
Notably, our model can be extended to multi-player and multi-strategy games.

\section{Materials and methods}\label{section:MM}

\subsection{Model with explicit time delays}
	
We begin by reintroducing the delay model constructed by Mi\c{e}kisz and Bodnar \cite{miekisz:PRE:2021}.
We will consider two-player symmetric games with two strategies:
cooperation (C) and defection (D) given by the following payoff matrix:
\begin{equation}\label{matrix:general}
\begin{array}{c}
\\
 C \\
 D 
\end{array}
\begin{array}{c}
\begin{matrix}
C & \!D 
\end{matrix} \\
\left(\ \begin{matrix}
R & S \\
T & P \\
\end{matrix}\ \right),
\end{array}
\end{equation}
where the \emph{i j} entry is the payoff of the first (row) player when it plays the strategy \emph{i} and the second (column) player plays the strategy \emph{j},  with $i,j \in \{C, D\}$. We assume that both players are
the same, and hence, payoffs of the column player are given by
the matrix transposed to \eqref{matrix:general}; such games are called symmetric.

Herein we follow \cite{miekisz:PRE:2021} closely.
Let us assume that during a time interval of length $\epsilon$,
only an $\epsilon$-fraction of the population takes part in pairwise competitions, that is, plays games.
Let $p_{i}(t)$  with  $i \in \{C, D\}$, be the number of individuals playing at time $t$ the strategy $C$
and $D$, respectively.
Then $p(t) = p_{C}(t) + p_{D}(t)$ is the total number of players and $x(t) = \frac{p_{C}(t)}{p(t)}$ is the fraction of the population playing $C$.
The expected payoffs of an individual is given by,
	\begin{equation}
	U_C (t) = R x(t) + S \left(1-x(t)\right),
	\end{equation}
and 
\begin{equation}
U_D (t) = T x(t) + P \left(1-x(t)\right),
\end{equation}
for cooperators and defectors, respectively.

We consider that individuals are born some time units after their parents played. 
We assume that time delays depend on strategies and are equal to $\tau_{C}$ and $\tau_{D}$ respectively.
 Mi\c{e}kisz and Bodnar \cite{miekisz:PRE:2021} proposed that the growth of the population playing a particular strategy is given as 
\begin{equation}
p_{i}(t + \epsilon) = (1-\epsilon)p_{i}(t) + \epsilon p_{i}(t-\tau_{i})U_{i}(t-\tau_{i}); i = C, D.
\end{equation}
and derive a system of coupled delayed differential equations for the fraction using the first strategy $x(t)$ and the population size $p(t)$,
	\begin{eqnarray}
\frac{dx(t)}{dt} &=& \frac{x(t-\tau_C)p(t-\tau_C)U_C(t-\tau_C)\left(1-x(t)\right)}{p(t)}\nonumber \\
&&-\frac{\left(1-x(t-\tau_D)\right)p(t-\tau_D)U_D(t-\tau_D)x(t)}{p(t)},\\
		\frac{dp(t)}{dt} &=& -p(t)+\left(p_C(t-\tau_C)U_C(t-\tau_C)\right)+p_D(t-\tau_D)U_D(t-\tau_D).
	\end{eqnarray}
 Then, the authors provided an equation for the stationary state of $x$ and solved it numerically for various games.
	
Here, we propose another type of model to deal with time delays.
We do not have to track the population size, and we are able to derive analytical formulas for the frequency of strategies in stationary states. 
This provides more insight into the dynamics of more complex games and can easily be expanded to multi-player and multi-strategy games.

\subsection{Kindergarten compartment model}
	
\begin{figure*}
		\centering
		\includegraphics[width=\textwidth]{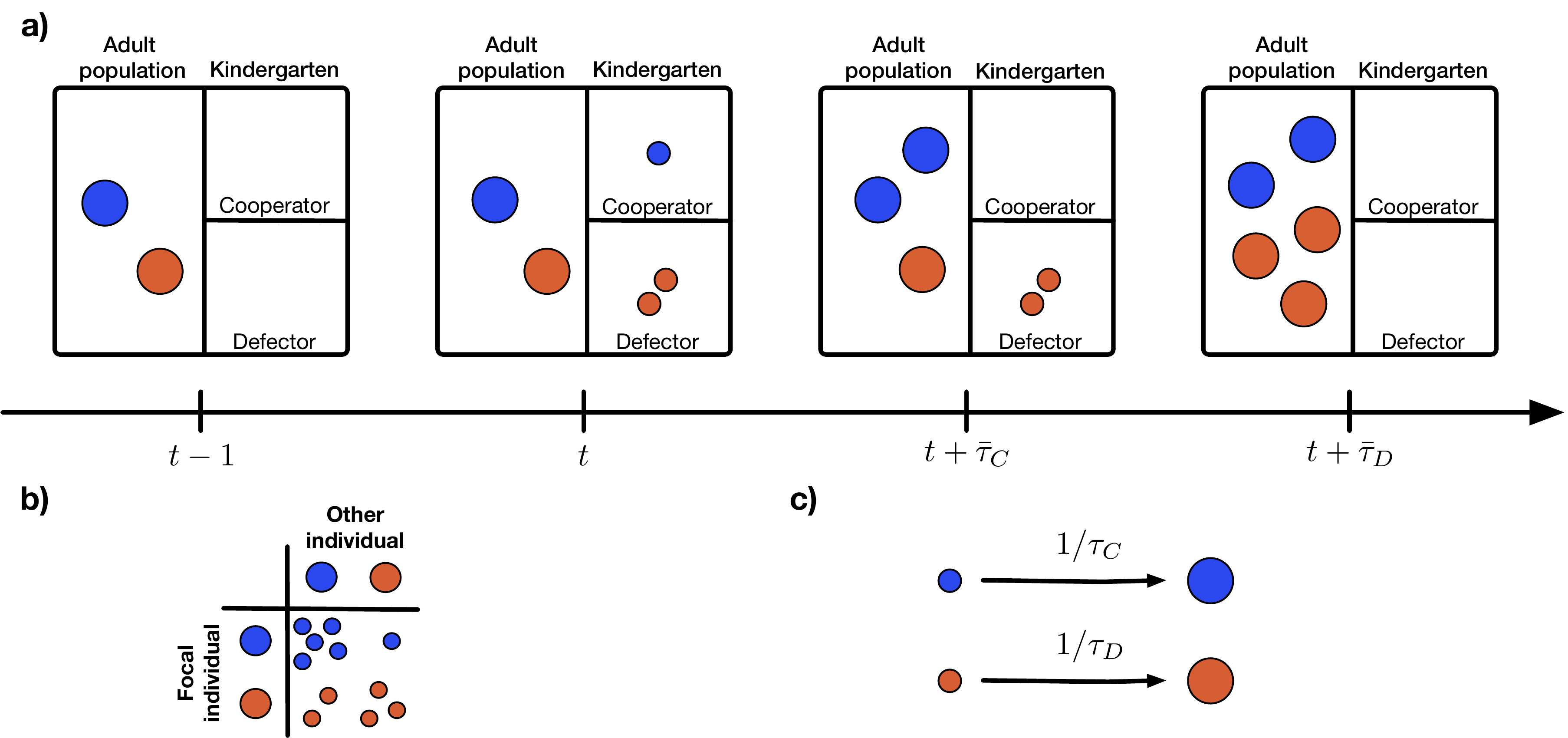}
		\caption{\textbf{a)} Individuals in the adult population interact with one another and receive payoff (offspring) based on those interactions. Offspring created at time $t$ join the strategy-specific kindergarten. After an offspring matures, it joins the adult population and can reproduce. The average maturation time is denoted by $\bar{\tau}_i$ for $i \in \{C, D\}$
\textbf{b)} The payoff matrix indicates the expected number of offspring created in each interaction.
\textbf{c)} Maturation rate depends on the parent's strategy and equals $1/\tau_C$ and $1/\tau_D$ for cooperators and defectors, respectively.}\label{fig:explain}
	\end{figure*}

We construct a model where time delays are represented by the inverse of rates at which an offspring grows and can participate in games. 
More precisely, the newly created offspring are located in inactive compartments. 
Then, with some strategy-dependent rates, they become players and move to the active compartment, as shown in Figure~\ref{fig:explain}.

Let us denote by $k_{i}$ sizes of kindergartens for offspring which inherited the $i$-th strategy with $i\in\{C, D\}$. 
Juveniles leave the kindergarten to join the adult population at a specific strategy-dependent rate.
Then, let $y_i= k_i /p$ for $i \in \{C, D\}$ denote sizes of kindergarten compartments relative to the size of the adult population.
Notably, in the kindergarten model, we relax the assumption made by~\cite{miekisz:PRE:2021} on the fraction of individuals partaking in the game in any small time interval, representing it instead by any non-decreasing function of the time passed, leading to a rescaling of time. The details of the function and the rescaling factor are shown in~\ref{modelderivation}.

Then, following~\ref{modelderivation}, we can derive the system of differential equations representing the change in the fraction of cooperators and the relative sizes of the kindergartens.
	\begin{align}\label{compartment}
		\frac{dx(t)}{dt} &= \frac{y_C(t)\left(1-x(t)\right)}{\tau_C}-\frac{y_D(t)x(t)}{\tau_D}, \nonumber \\
\frac{dy_C(t)}{dt} &= y_C(t) \left(\frac{\tau_C-1}{\tau_C}-\frac{y_C(t)}{\tau_C}-\frac{y_D (t)}{\tau_D}\right)+x(t)U_C(t), \\
\frac{dy_D(t)}{dt} &= y_D(t)\left(\frac{\tau_D-1}{\tau_D}-\frac{y_C(t)}{\tau_C}-\frac{y_D (t)}{\tau_D}\right)+\left(1-x(t)\right)U_D(t). \nonumber
	\end{align}
Similarly, we can obtain the differential equation for the change in population size, given by:
\begin{align}
\frac{dp(t)}{dt} &= p(t) \left(\frac{y_C (t)}{\tau_C}+\frac{y_D (t)}{\tau_D} - 1\right).
	\end{align}
Hence, the size of the population in a stationary state of frequencies $x^*, y_C^*, y_D^* $, $p(t) = p(0) e ^{(y_C^*/\tau_C + y_D^*/\tau_D -1)t}$ grows exponentially if 
\begin{equation}\label{lambdastar}
	\rho(\tau_C,\tau_D)=\frac{y_C^*(t)}{\tau_C}+\frac{y_D^*(t)}{\tau_D} >1,
\end{equation}
or goes extinct otherwise.
Hence, condition~\eqref{lambdastar} has to be met for the population not to go extinct.
We should take care of cases in which one delay is equal to zero. 
For $\tau_C = 0$ there is no cooperator kindergarten ($y_C = 0$) and the system~\eqref{compartment} becomes:
\begin{align}\label{compartmenttc0}
	\frac{dx(t)}{dt} &= x(t)\left(U_C (t) (1-x(t))-\frac{y_D(t)}{\tau_D}\right), \nonumber \\
\frac{dy_C(t)}{dt} &= 0, \\
\frac{dy_D(t)}{dt} &= y_D(t) \left(\frac{\tau_D-1}{\tau_D}-x(t)U_C(t)-\frac{y_D (t)}{\tau_D}\right)+\left(1-x(t)\right)U_D(t). \nonumber
\end{align}
Analogously, for $\tau_D = 0$ the defector kindergarten compartment is empty ($y_D = 0$) and the system~\eqref{compartment} becomes:
\begin{align}\label{compartmenttd0}
	\frac{dx(t)}{dt} &= \frac{y_C(t)}{\tau_C} - x(t)\left((1-x(t))U_D(t)+\frac{y_C(t)}{\tau_C}\right), \nonumber \\
\frac{dy_C(t)}{dt} &= y_C(t) \left(\frac{\tau_C-1}{\tau_C}-(1-x(t))U_D(t)-\frac{y_C (t)}{\tau_C}\right)+x(t)U_C(t), \\
\frac{dy_D(t)}{dt} &= 0. \nonumber
\end{align}	
The system of equations~\eqref{compartment} has two trivial stationary states, $x=1, 0$, of full cooperation and full defection, and two possible internal stationary states whose existence depends on the game parameters and delays.

For strategy-independent delays, i.e. $\tau_C = \tau_D$, our equations simplify greatly, and the system has only one possible internal stationary state:
$x^*= \left(P-S \right) / \left(R-T+P-S\right)$
equal to the one for replicator dynamics without time delays. 
Hence, in this work we only focus on strategy-dependent delays, i.e. $\tau_C = \tau_D$.


In the following, we study Eqs.~\eqref{compartment} through the lenses of dynamical systems and bifurcation theory.
That is, we investigate stationary states of Eqs.~\eqref{compartment} and perform a linear stability analysis of their stability ~\cite{glendinning1994stability}.
They might be given in a closed form for special payoff matrices, allowing us to analyse the dependence of the stationary states on delays and payoff matrix entries.
In the case of a general payoff matrix, stationary states are calculated numerically (Julia package BifurcationKit~\cite{veltz:hal-02902346}).
In particular, we consider three specific games, Stag-Hunt, Snowdrift, and the Prisoner's Dilemma one.

\section{Results}\label{results}
\subsection{Stag-Hunt game}
In the Stag-Hunt game, two individuals decide between cooperating and hunting a stag together (C) or pursuing a hare independently (D).
Hunting a hare does not require help from the other individual and results in a specific payoff $b$; however, a stag provides a higher payoff value $a$ \cite{skyrms:book:2004}. 
The following payoff matrix characterises the Stag-Hunt game:
\begin{equation}
\begin{array}{c}
\\
 C \\
 D 
\end{array}
\begin{array}{c}
\begin{matrix}
C & \! & \!& \! D 
\end{matrix} \\
\left(\ \begin{matrix}
R=a & S=0 \\
T=b & P=b \\
\end{matrix}\ \right)
\end{array}
\end{equation}
where $a>b>0$.
In the case of no delays, i.e. $\tau_D=\tau_C=0$, Eq.~\eqref{compartment} collapses to a well-studied replicator dynamics with two asymptotically stable stationary states, $x_0=0$ and $x_1=1$, and an unstable one, $x_2 = b/a$.
For the Stag-Hunt game with compartments, the system~\eqref{compartment} becomes:
\begin{align}\label{compartmentxsh}
		\frac{dx(t)}{dt} &= \frac{y_C(t)(1-x(t))}{\tau_C}-\frac{y_D(t)x(t)}{\tau_D}, \nonumber \\
		\frac{dy_C(t)}{dt} &= y_C \left(\frac{\tau_C-1}{\tau_C}-\frac{y_C(t)}{\tau_C}-\frac{y_D (t)}{\tau_D}\right)+a x^2(t), \\
		\frac{dy_D(t)}{dt} &= y_D\left(\frac{\tau_D-1}{\tau_D}-\frac{y_C(t)}{\tau_C}-\frac{y_D (t)}{\tau_D}\right)+b (1-x(t)). \nonumber 
\end{align}
The system \eqref{compartmentxsh} may have three stationary states,  in $[0, 1]^3$.  
The full defection stationary state $x_0=0$ becomes 
\begin{equation}
e_0=\left(0, 0, \frac{1}{2} \left(\tau_D-1+\sqrt{4 b \tau_D+\tau_D^2-2 \tau_D+1}\right)\right).
\end{equation}
For the condition~\eqref{lambdastar} to be met and the population in this stationary state to grow exponentially, the following needs to be true: 
 \begin{equation}
\frac{1}{2 \tau_D} \left(\tau_D-1+\sqrt{4 b \tau_D+\tau_D^2-2 \tau_D+1}\right)>1.
 \end{equation}
Thus, an additional condition needs to hold: $b \geq 1$.
 The full cooperation stationary state $x_1$ becomes
\begin{equation}
e_1=\left(1, \frac{1}{2} \left(+\tau_C-1+\sqrt{4 a \tau_C+\tau_C^2-2 \tau_C+1}\right), 0\right) .
\end{equation}
Again, we check the condition~\eqref{lambdastar}, which in this stationary state becomes:
\begin{equation}
\frac{1}{2\tau_C} \left(\tau_C-1+\sqrt{4 a \tau_C+\tau_C^2-2 \tau_C+1}\right)>1.
\end{equation}
For the condition to held, we need an additional constraint: $a \geq 1$.
Lastly, the internal stationary state $x_2$ takes the following form:
\begin{equation}
e_2 = \left(x_2^*, -\frac{\tau_C x_2^* y_D^*}{\tau_D (x_2^*-1)}, y_D^*\right) 
\end{equation}
where
\begin{align}
	x^*_2 = &\frac{\tau_C+(\tau_D-\tau_C)\sqrt{(4 b-2) \tau_D+\tau_D^2+1}}{2 a \tau_D^2}\nonumber \\ 
&+\frac{2 b \tau_C -\tau_C +\tau_D-1}{2 a \tau_D},\\
y^*_D = &-\frac{(x-1) \tau _D \left(\tau _C-1+\sqrt{(4 a x-2) \tau _C+\tau _C^2+1}\right)}{2 \tau _C}.
	\end{align}
For the condition~\eqref{lambdastar} to be met in $e_2$, that is for
\begin{equation}
-\frac{\tau_C x_2^* y_D^*}{\tau_D \tau_C (x_2^*-1)} + \frac{y_D^*}{\tau_D} >1
\end{equation}
to hold true, we need $x_2^* \geq 1/a$.
For the population to grow exponentially we have to have $a, b>1$ and $x_2^* \geq 1/a$.

\textit{Homogeneous stationary states.}
We conduct stability analysis of $e_0$ and show that it is always a stable stationary state.
The stationary state $e_1$ is stable unless the following condition is met: $\tau_C\geq m$ where $$m = \frac{b \left(b \sqrt{\frac{(a-b)^2 \left(4 b \tau_D+\tau_D^2-2 \tau_D+1\right)}{(b-1)^2 b^2}}-\sqrt{\frac{(a-b)^2 \left(4 b \tau_D+\tau_D^2-2 \tau_D+1\right)}{(b-1)^2 b^2}}-\tau_D-1\right)+a (2 b-1) \tau_D+a}{2 (b-1) b}.$$ 
For a large enough cooperator delay, full cooperation loses its stability, the internal stationary states vanishes.
Details of the analysis are presented below and in~\ref{SA_SH}.

 \textit{Internal stationary state.}
If no delays are present, the internal stationary point $x_2^*$ always exists and equals to $b/a$.
The increase of $\tau_C$ leads to an increase in the stationary point value.
Then, the stationary state reaches full cooperation ($x_2^*=1$) and disappears when $\tau_C = m$.
An increase in $\tau_D$ always leads to a decrease in $x^*_2$.  
In the limiting case of $\tau_D \to \infty$\ we have  $x_2^* \to 1/a$. 
The latter result ensures that condition~\eqref{lambdastar} is met for the internal stationary state.
When the internal stationary state $e_2$ exists, it is always unstable, as shown in~\ref{SA_SH}.

  \textit{One delay present.}
We consider the limiting case of each delay to be equal to 0. 
We solve the systems \eqref{compartmenttc0} for $\tau_C=0$ and \eqref{compartmenttd0} for $\tau_D=0$ and check the behaviour of $e_2$ in the corresponding limits. If it exists, the internal stationary state of the system \eqref{compartmenttc0} takes the following form:
\begin{equation}
\tilde{e}_2 = \left(\tilde{x}_2, 0, -a \tau_D (\tilde{x}_2-1) \tilde{x}_2\right)
\end{equation}
where
\begin{equation}
\tilde{x}_2=\frac{\tau_D-1+\sqrt{4 b \tau_D+\tau_D^2-2 \tau_D+1}}{2 a \tau_D}.
\end{equation}
 
	The solution coincides with the value of  $e_2$ in the limit of $\tau_C \to 0$. Additionally, we can show that: $ \lim_{\tau_D \to 0} \tilde{x}_2 = b/a $. 
Hence, in the limiting case of no delays, we recover the solution of the standard Stag-Hunt game. 
	Similarly, for $\tau_D=0$, we solve the system~\eqref{compartmenttd0}. The system has only one stationary solution in the  game-theoretic relevant interval of $\bar{x} \in(0, 1)$:
  \begin{equation}
 \bar{e}_2 = (\bar{x}_2, b \tau_C \bar{x}_2, 0)
  \end{equation}
where
\begin{equation}
\bar{x}_2=\frac{b (b \tau_C-\tau_C+1)}{a}.
\end{equation}
	 
	 Again, we can show that $\lim_{\tau_D \to 0} x_2^* = \bar{x}_2$ and $\lim_{\tau_C \to 0}\bar{x}_3 = b/a $. So, the internal solution represented by $e_2$ can be used in the limiting cases of either or both delays approaching 0. 
	 
	 We show that in the classical Stag-Hunt game, the introduction of delay is insufficient to destabilise complete defection. However, it is possible to destabilise full cooperation. Moreover, we prove that increasing delay of one of the strategies leads to shrinking of the basin of attraction of the stationary state consisting of individuals following only that strategy.

\textit{Example.}
Below, we present an analysis of the Stag-Hunt game characterized by the following payoff matrix:
	\begin{equation}\label{matrix:staghunt}
\begin{array}{c}
\\
C \\
D 
\end{array}
\begin{array}{c}
\begin{matrix}
C & \! & \!& \! D 
\end{matrix} \\
\left(\ \begin{matrix}
R=5 & S=0 \\
T=3 & P=3 \\
\end{matrix}\ \right)
\end{array}
\end{equation}
	
In the case of no delay, the game has two stable stationary states corresponding to the two homogeneous states. The unstable stationary state, dividing the basins of attraction of the two stationary states, is $x^* = 0.6$.
	
	\begin{figure*}
\centering
\includegraphics[ width=0.9 \textwidth]{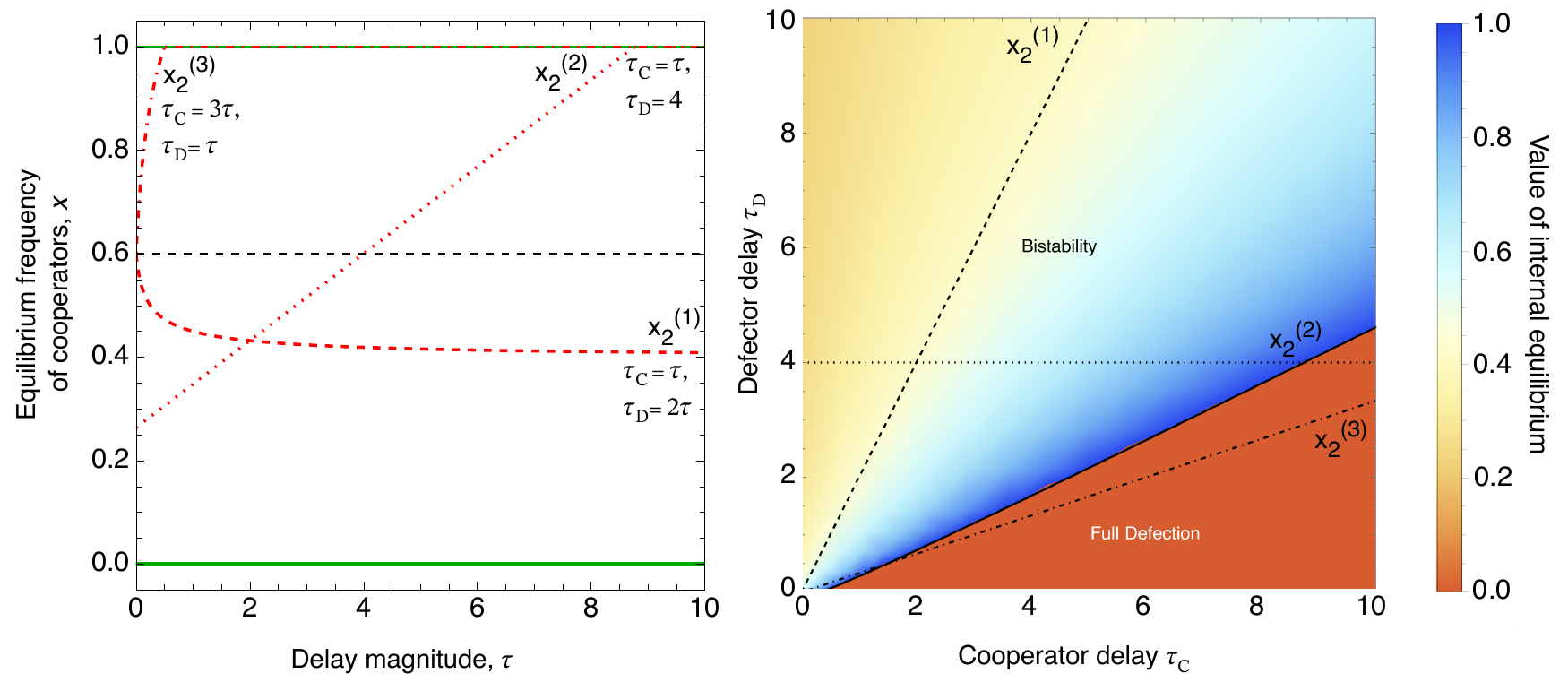}
\caption{Stability of the stationary states
of the Stag-Hunt game represented by matrix~\eqref{matrix:staghunt}. 
On the left, the stationary state values for the specific values of delays are plotted.
The dashed line represents the internal stationary state $x_2^{(1)}$ as a function of $\tau$, when $\tau_C = \tau$, $\tau_D = 2 \tau$. An increase in $\tau$ leads to a decrease in stationary state value.
The dotted line represents internal stationary state $x_2^{(2)}$ as a function of $\tau$, when $\tau_C = \tau$, $\tau_D = 4$. The internal stationary state disappears for big enough $\tau$, and full cooperation loses its stability.
The dot-dashed line represents the internal stationary state $x_2^{(3)}$ as a function of $\tau$, when $\tau_D = \tau$, $\tau_C = 3 \tau$. An increase in $\tau$ leads to an increase in the stationary state value until its disappearance. 
On the right, the stability of the system in the parameter space $\tau_C$ and $\tau_D$ is shown. 
The solid black line indicates the point of bifurcation. In the "Full Defection" region, only $e_0$ is stable. In the "Bistability" region, both $e_0$ and $e_1$ are stable, and their basins of attraction are divided by the unstable internal stationary state. The colour indicates the value of the internal stationary state. The values considered on the left are represented by dashed ($x_2^{(1)}$), dotted ($x_2^{(2)}$) and dot-dashed ($x_2^{(3)}$) lines, respectively.
The effects of only one delay can be observed on the left and bottom edges of the plot.}
\label{fig:SH}
\end{figure*}
Figure \ref{fig:SH} explores the change in the stability of the stationary states of the system~\eqref{compartmentxsh} in the parameter space of the delays. 
In most of the parameter space, we observe full cooperation and defection bistability. 
The value of $ x^*_2$ decreases with $\tau_D$, reaching the limiting value of $0.2$ ($1/a$). 
With the increase in $\tau_C$, an increase in the value of the $x^*_2$ is observed. 
As the internal stationary state value approaches $1$, one of the stable stationary states disappears, leaving full defection as the only stable stationary state. 
The curve dividing the region of bistability and full defection marks the bifurcation point.

\subsection{Snowdrift game}
Next, we analyse the Snowdrift game, also known as the chicken game or the hawk-dove game. In this game, two players choose between contributing to a common good (C) or not (D). The cost of the good ($c$) is divided equally between all contributors. If at least one of the individuals contributes, each player, regardless of their strategy, obtains the benefit $b$ \cite{osborne:book:2004}. Hence, the game can be represented by the following matrix: 
\begin{equation}\label{matrixgenSD}
\begin{array}{c}\\
C \\
D 
\end{array}
\begin{array}{c}
\begin{matrix}
C & \! & \!& \! &\!& \!D 
\end{matrix} \\
\left(\ \begin{matrix}
R=b-c/2 & S=b-c \\
T=b & P=0 \\
\end{matrix}\ \right)
\end{array}
\end{equation}
where $b>c>0$.
The game with no delays has two unstable stationary states $x_0 = 0$, $x_1 = 1$ and one stable internal stationary state $x_2 = (b-c)/(b-c/2)$. 

For the payoffs specified in the matrix~\eqref{matrixgenSD} the system~\eqref{compartment} becomes:
\begin{align}\label{compartmentsg}
\frac{dx(t)}{dt} =& \frac{y_C(t)(1-x(t))}{\tau_C}-\frac{y_D(t)x(t)}{\tau_D} \nonumber \\
\frac{dy_C(t)}{dt} =& y_C \left(\frac{\tau_C-1}{\tau_C}-\frac{y_C(t)}{\tau_C}-\frac{y_D (t)}{\tau_D}\right)+x(t)\left(\frac{c}{2}x(t) +b -c\right)\\
\frac{dy_D(t)}{dt} =& y_D\left(\frac{\tau_D-1}{\tau_D}-\frac{y_C(t)}{\tau_C}-\frac{y_D (t)}{\tau_D}\right)+b x(t) (1-x(t)). \nonumber
\end{align}
This system~\eqref{compartmentsg} can have four stationary states. 
The trivial stationary state $x_0$ becomes:
\begin{equation}
e_0 = \left(0, 0,\tau_D-1 \right),
\end{equation}
if $\tau_D>1$. 
The population would grow exponentially if $1/\tau_D <0$. 
Additionally, when $\tau_D \leq 1$, $y_D = 0$.
As the delay cannot take negative values, we can see that the population always goes extinct in the full defection stationary state. 

The full cooperation stationary state $x_1$ becomes:

\begin{equation}
e_1 = \left(1, \frac{1}{2} \left(\tau_C-1+\sqrt{\tau_C (4 b-2 c+\tau_C-2)+1}\right), 0\right).
\end{equation}

In the full cooperation stationary state, the condition~\eqref{lambdastar} becomes: 
\begin{equation}\label{lambdacoopsg}
\frac{1}{2\tau_C} \left(\tau_C-1+\sqrt{\tau_C (4 b-2 c+\tau_C-2)+1}\right)>1
\end{equation}
and holds when $(1<b<2\land 0<c\leq 2 b-2)\lor (b\geq 2)$.
Additionally, two internal stationary states may be present:
\begin{align}
e_2 =& \left(x_2^*,\frac{\tau_C x_2^* y_D^{(2)}}{\tau_D-\tau_D x_2^*}, y_D^{(2)}\right) \\
e_3 =& \left(x_3^*,\frac{\tau_C x_3^* y_D^{(1)}}{\tau_D-\tau_D x_3^*}, y_D^{(3)}\right)
\end{align}
where
\begin{align}
x^*_2 =& \frac{1}{(c \tau_D-2 b \tau_C)^2} \nonumber \\
&\Big((\tau_C-\tau_D)  \sqrt{o}  \nonumber \\
& +\tau_D \left(4 b^2 \tau_C-2 b (2 c \tau_C+\tau_C-1)-c (\tau_C+1)\right) \nonumber \\
&+\tau_C (2 b (\tau_C-1)+c)+c \tau_D^2 (-2 b+2 c+1)\Big), \label{x2defo}\\
x^*_3 = & \frac{1}{(c \tau_D-2 b \tau_C)^2} \nonumber \\
&\Big((\tau_D-\tau_C)  \sqrt{o}  \nonumber \\
& +\tau_D \left(4 b^2 \tau_C-2 b (2 c \tau_C+\tau_C-1)-c (\tau_C+1)\right) \nonumber \\
&+\tau_C (2 b (\tau_C-1)+c)+c \tau_D^2 (-2 b+2 c+1)\Big), \label{x3defo}\\
y_d^{(i)} =& -\frac{1}{2} (x^*_i-1) \left(\tau_D-1+\sqrt{4 b \tau_D x+\tau_D^2-2 \tau_D+1}\right).\\
\end{align}

In~\eqref{x2defo} and~\eqref{x3defo} $o$ is defined as: 

\begin{align}
    o=&16 b^3 \tau_C+4 b^2 (\tau_C (-4 c+\tau_C-2)-2 c \tau_D+1)  \nonumber \\
&+4 b c (2 c \tau_D+\tau_C (-\tau_D)+\tau_C+\tau_D-1)+c^2 (\tau_D-1)^2. \label{o_eq} 
\end{align}
In the internal equilibira, the condition~\eqref{lambdastar} takes the following form: 
\begin{equation}
\frac{ x_i^* y_D^{(i)}}{\tau_D-\tau_D x_i^*}+\frac{y_D^{(i)}}{\tau_D} >1.
\end{equation}
The condition is met when one of the following constraints holds true: 
\begin{align}\label{lambdainternalsg}
&\biggl(1<b\leq 2\land \biggl((0<c\leq b-1) \nonumber\\
&\lor\left(b-1<c<2 b-2\land \frac{-2 b+2 c+2}{c}\leq x_i^*<1\right)\biggl)\biggl),\\
&\biggl(b>2\land \biggl((0<c\leq b-1) \nonumber \\
&\lor\left(b-1<c<b\land \frac{-2 b+2 c+2}{c}\leq x_i^*<1\right)\biggl)\biggl). \nonumber
\end{align}

Now, we combine conditions~\eqref{lambdacoopsg} and~\eqref{lambdainternalsg} and obtain the following constrains on the game parameters: $(1<b \leq 2 \land  c < 2b-2 ) \lor b>2 $.
Additionally, when $b<c+1$, the following constraint needs to be applied to the internal stationary state values: $\frac{-2 b+2 c+2}{c}\leq x_i^*$ for $i \in \{2, 3\}$.

\textit{Homogeneous stationary states.}
The analysis of the full cooperation stationary state $e_1$ shows that it is stable if $\tau_D>n$ where $n=\frac{4 b^2 \tau_C-2 b c \tau_C-4 b \tau_C+c \tau_C+c}{4 b^2-4 b c-4 b+c^2+2 c}+\sqrt{\frac{4 b c^2 \tau_C-2 c^3 \tau_C+c^2 \tau_C^2-2 c^2 \tau_C+c^2}{\left(4 b^2-4 b c-4 b+c^2+2 c\right)^2}}$. Hence, full cooperation can be stable if the defector delay is large enough and the cooperator delay is small enough.
We show that full defection stationary state $e_0$ is stable if the following conditions are met $\left( 0<c\leq 2 \land (2+c)/c <b<c+1\land \tau_D>p \right)\lor \left(c>2 \land c< b<c+1 \land \tau_D>p\right)$ where $p=\frac{1}{2} \sqrt{\frac{4 b \tau_C-4 c \tau_C+\tau_C^2-2 \tau_C+1}{(b-c-1)^2}}+\frac{-\tau_C-1}{2 (b-c-1)}$ and is never stable if $\tau_D <1$. 
The specific form of the payoff matrix of the Snowdrift game violates the assumption of exponential growth whenever full defection is in a stable stationary state. 
Therefore, any parameter combination leading to stability of full defection needs to be noted as leading to the possible extinction of the population. 
As population extinction is excluded from the model's assumptions, we only analyse the parameter space when it is not possible. 
Hence, we assume that $b \geq c+1$.
The details of the stability analysis are presented in \ref{SA_SG}.

 \textit{Internal stationary state.}
We analyse the two possible internal (in the interval $(0, 1)$) stationary states ($e_2$, $e_3$).
When the cooperator delay is greater than the defector delay ($\tau_C>\tau_D$), only one internal stationary state, $e_2$, exists. 
In that region of the parameter space, the stationary state is always stable. 
With the increase of cooperator delay, the internal stationary state value $x_2^*$  decreases. In the limit of $\tau_C \to \infty$ $x_2^*$ approaches a limiting value $1/b$. Notably, for all considered game parameters $1/b>(-2 b+2 c+2)/c$, hence the population always grows exponentially in the internal stationary state.
An increase in the defector delay $\tau_D$ leads to an increase in the value of $x_2^*$. 
In the limit of the delay $\tau_D$ approaching $\tau_C$, the stationary state approaches the limiting value, the internal stationary state of the system with no delays, $(b-c)/(b-c/2)$. 

\begin{figure}
\centering
\includegraphics[width=0.5\columnwidth]{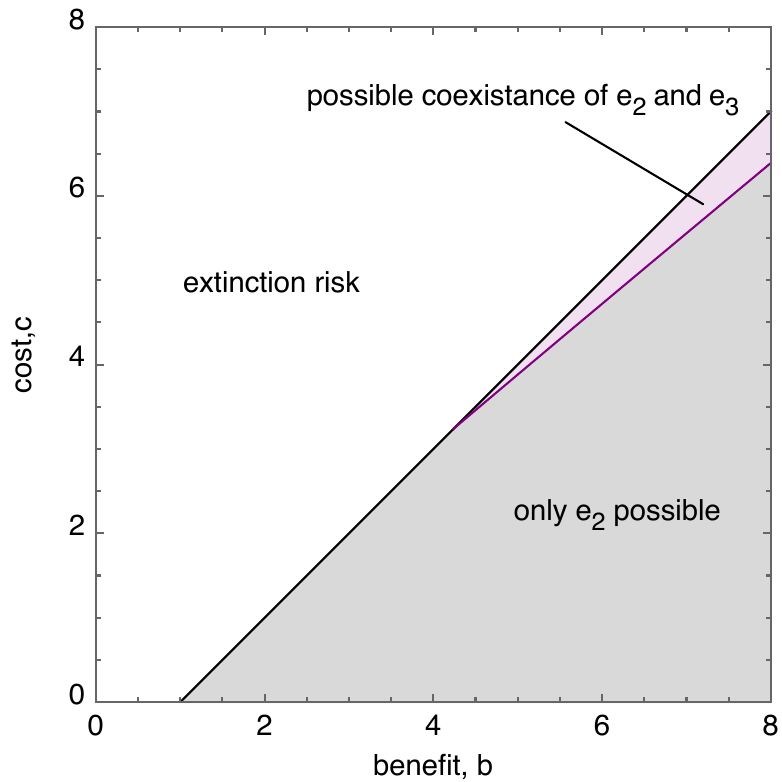}
\caption{The existence of one or two internal in the Snowdrift game parameter ($b$ and $c$) space. In the grey region, only $e_2$ can exist. In the purple region, both $e_2$ and $e_3$ may exist. Outside of the coloured region, the population is at risk of extinction.}
\label{fig:SG_params}
\end{figure}

If the defector delay is greater than the cooperator delay ($\tau_D>\tau_C$), both internal stationary states may exist in the interesting interval $(0, 1)$. 
In particular, the existence of the stationary state $e_3$ depends on the parameter values of the game: the stationary state can exist only if certain conditions are met. Otherwise, only $e_2$ can exist. 
The two regions of the $b, c$ parameter space are represented in Figure~\ref{fig:SG_params}. 
Both internal stationary states may exist in the parameter region indicated in purple in Fig.~\ref{fig:SG_params}.
Particularly, when $0<\tau_C<q$, where $q=\frac{8 b^3-4 b^2 (3 c+2)+4 b c (c+2)-c^2}{2 \sqrt{2} \sqrt{b \left(2 b^3-2 b^2 (2 c+1)+b c (2 c+3)-c^2\right)}}-2 b+c+1$ the two stationary states coexist, with $x_2^*<x_3^*$, $e_2$ being stable and $e_3$ unstable. 
At the point $o=0$, where $o$ has been defined as~\eqref{o_eq}, the two stationary states collide at a fold bifurcation and disappear. 
$e_3$ collides with the full cooperation stationary state $e_1$ at $\tau_D=n$ and the two switch stability. 
In the gray region in Fig.~\ref{fig:SG_params} or when $\tau_C \geq q$ only $e_2$ exists. The value of $x_2^*$ decreases with cooperator delay $\tau_C$ and increases with defector delay $\tau_D$, colliding with the full cooperation stationary state at $\tau_D=n$. In the limit of $\tau_C \to \tau_D$ the fraction of the cooperators approaches $(b-c)/(b-c/2)$. If it exists, the internal stationary state is always stable. 
 
 \textit{One delay present.}
To analyse the system~\eqref{compartmenttc0}, we must consider the game parameter values and the two regions in Fig.~\ref{fig:SG_params}. 
In the grey region only one internal stationary state can exist, the fraction of cooperators coinciding with $x_2^*$ in the limit of $\tau_C \to 0$:
\begin{align}\label{x2tildasnow}
\tilde{x}_2 =& \frac{1}{c^2 \tau_D}\Big(b (2-2 c \tau_D)+2 c^2 \tau_D+c \tau_D-c \nonumber\\
& -\sqrt{b^2 (4-8 c \tau_D)+4 b c (2 c \tau_D+\tau_D-1)+c^2 (\tau_D-1)^2}\Big) \\
=& \lim_{\tau_C \to 0} x_2^*. \nonumber
\end{align}
In the limit of defector delay tending to $0$, the internal stationary state approaches $x_2 = (b-c)/(b-c/2)$. 
In the purple region of Fig.~\ref{fig:SG_params} two internal stationary states can be present, corresponding to $e_2$ and $e_3$ such that $\tilde{x}_2$ is given by Eq.~\eqref{x2tildasnow} and $\tilde{x}_3$ by:
\begin{align}
 \tilde{x}_3 =& \frac{1}{c^2 \tau_D}\Big(b (2-2 c \tau_D)+2 c^2 \tau_D+c \tau_D-c \nonumber\\
& +\sqrt{b^2 (4-8 c \tau_D)+4 b c (2 c \tau_D+\tau_D-1)+c^2 (\tau_D-1)^2}\Big) \\
=& \lim_{\tau_C \to 0} x_2^*. \nonumber
\end{align}
In the limit of no delays present, only one of the internal stationary states exists in the interval $(0, 1)$, in particular, $\lim_{\tau_D \to 0}  \tilde{x}_2  = (b-c)/(b-c/2)$.

If only cooperators experience delays ($\tau_D = 0$) the system~\eqref{compartmenttd0} has one possible internal stationary state $\bar{e}_2$. The value of $\bar{x}_2$ is equal to the stationary state value $x_2^*$ in the limit of no defector delays:
\begin{align}
 \bar{x}_2 =& \frac{1}{4 b^2 \tau_C}\Big(2 b \tau_C-2 b+c \nonumber \\
& +\sqrt{8 b^2 \tau_C (2 b-2 c)+(2 b \tau_C-2 b+c)^2}\Big) \\
=&  \lim_{\tau_D \to 0} x_2^*. \nonumber
\end{align}
With the cooperator delay approaching 0, the value of the internal stationary state approaches the no-delay value: $\lim_{\tau_C \to 0} \bar{x}_2 = x_2 = (b-c)/(b-c/2)$.
 
\textit{Examples.}
We perform numerical analysis on two payoff matrices, each corresponding to one region in Figure \ref{fig:SG_params}. Matrix~\eqref{matrix:snow1} ($b=5$ and $c=2$) corresponds to the grey region and the matrix~\eqref{matrix:snow2} ($b=5$, $c=3.9$) lies in the purple region.
\begin{equation}\label{matrix:snow1}
\begin{array}{c}
\\
 C \\
 D 
\end{array}
\begin{array}{c}
\begin{matrix}
C & \! & \!& \! & \!D 
\end{matrix} \\
\left(\ \begin{matrix}
R=4 & S=3 \\
T=5 & P=0 \\
\end{matrix}\ \right)
\end{array}
\end{equation}
In the case of no delays, the game characterized by the payoff matrix~\eqref{matrix:snow1} has one stable stationary state $x=0.75$ and two unstable stationary states of full cooperation and full defection.
\begin{figure*}
\centering
\includegraphics[width=0.9 \textwidth]{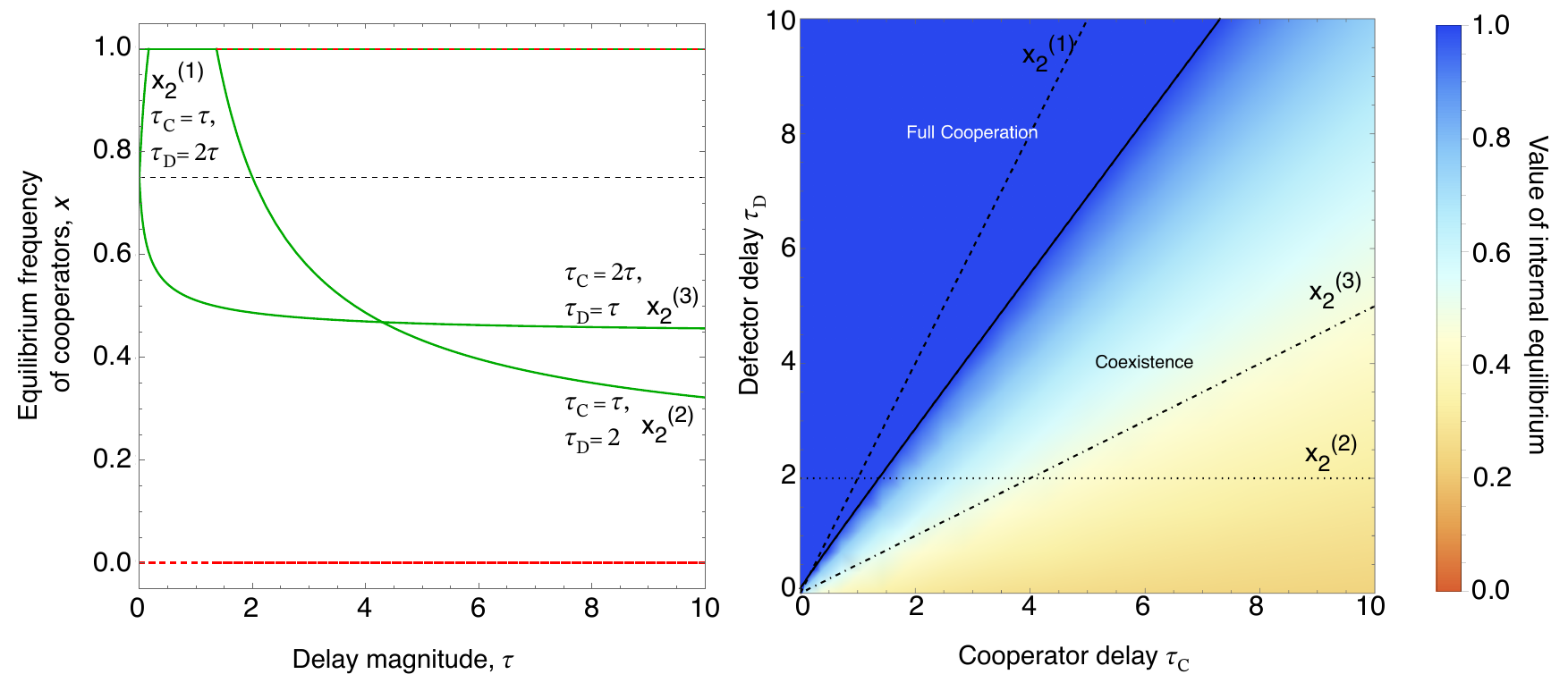}
\caption{Stability of the stationary states of the Snowdrift game represented by matrix~\eqref{matrix:snow1}. 
On the left, the stationary state values for specific delays are plotted.
The $x_2^{(1)}$  line represents the internal stationary state as a function of $\tau$, when $\tau_C = \tau$, $\tau_D = 2 \tau$. An increase in $\tau$ leads to an increase in stationary state value. The internal stationary state ceases to exist for a high enough value of $\tau$, and full cooperation becomes stable.
The $x_2^{(2)}$ line represents internal stationary state as a function of $\tau$,when $\tau_C = \tau, \tau_D = 2$. 
Full cooperation is a stable stationary state for small values of $\tau$, and no internal stationary state exists. 
A stable internal stationary state appears with the increase in $\tau$, and full cooperation loses stability. 
The value of the internal stationary state decreases with the increase of $\tau$.
The $x_2^{(3)}$ line represents the internal stationary state as a function of $\tau$ when $\tau_C = 2 \tau$, $\tau_D=\tau$. The internal stationary state's value decreases with an increase in $\tau$.
On the right, the stability of the system in the parameter space $\tau_C$ and $\tau_D$ is shown. 
The solid black line indicates the point of bifurcation. 
In the "Full Cooperation" region, only $e_1$ is stable. 
In the "Coexistence" region, only $e_2$ is stable, the fraction of cooperators indicated by the colour. 
The values considered on the left are represented by dashed ($x_2^{(1)}$), dotted ($x_2^{(2)}$) and dot-dashed ($x_2^{(3)}$) lines, respectively.
The effects of only one delay present can be observed on the left and bottom edges of the plot.}
\label{fig:SD1}
\end{figure*}

In Figure \ref{fig:SD1}, the behaviour of the internal stationary state is analysed. In the majority of the space, the internal stationary state exists and is stable. The internal stationary state disappears above the boundary value of $\tau_D$, and full cooperation becomes stable.
To showcase the dynamics in the purple region of Fig.~\ref{fig:SG_params}, we look at another payoff matrix:
\begin{equation}\label{matrix:snow2}
\begin{array}{c}
\\
 C \\
 D 
\end{array}
\begin{array}{c}
\begin{matrix}
C & \! & \!& \! & \!D 
\end{matrix} \\
\left(\ \begin{matrix}
R=3.05 & S=1.1 \\
T=5 & P=0 \\
\end{matrix}\ \right)
\end{array}
\end{equation}
\begin{figure*}
\centering
\includegraphics[width=0.9\textwidth]{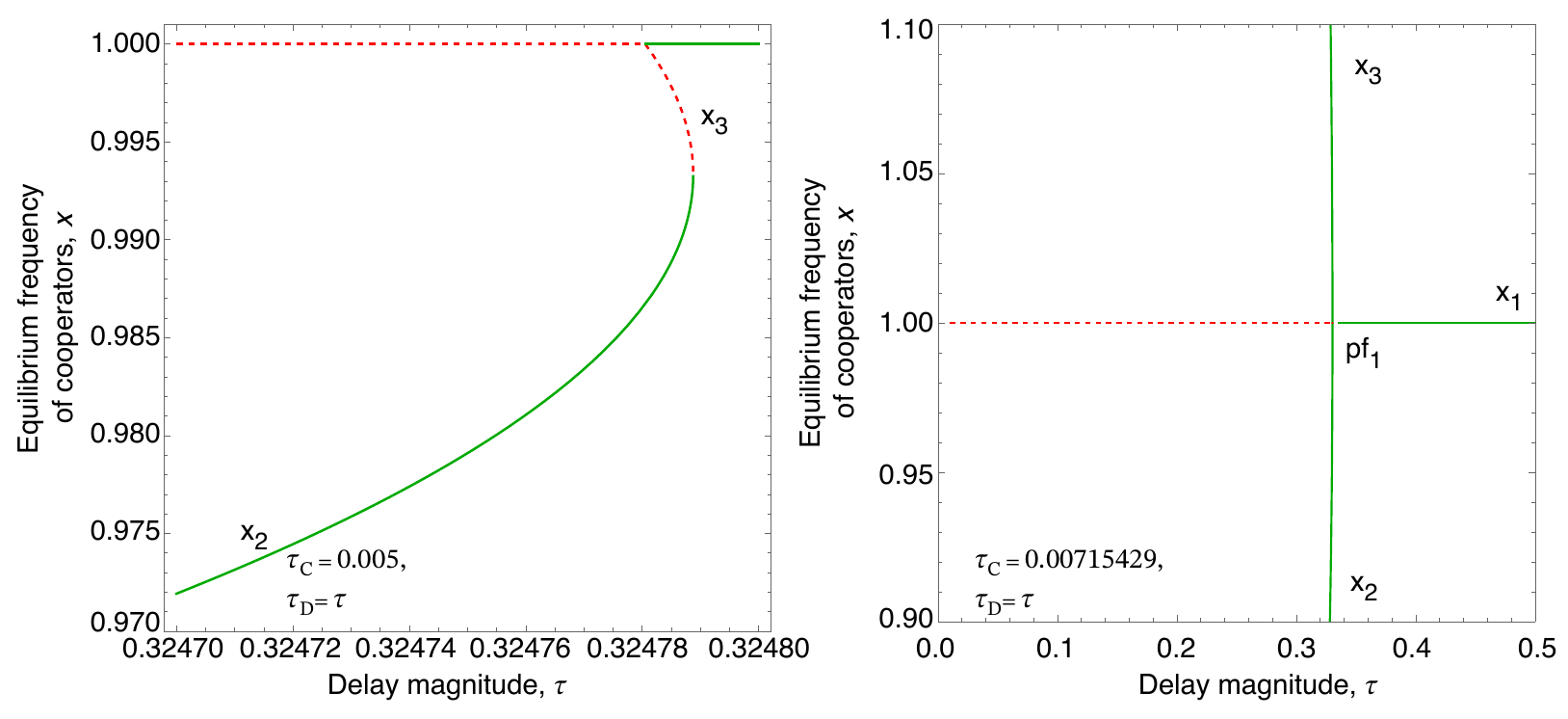}
\caption{The stability and the existence of internal stationary states of the Snowdrift game represented by matrix~\eqref{matrix:snow2} depend on the time delays.  
The left plot represents the internal stationary states' values as a function of $\tau$, when $\tau_C = 0.005$, $\tau_D =  \tau$. Two internal stationary states may exist at the same time. The value of $x_3^*$ decreases with $\tau$ and $x_2^*$  increases. $e_3$ is always unstable and appears when full cooperation becomes stable. After the internal stationary states collide, they disappear, and full cooperation is the only stable stationary state. Full defection is always unstable.
As shown on the right plot,  the two internal stationary states emerge at $\tau_C=0.00715429$, and $\tau_D\approx0.33031257$ as the saddle-node that gives rise to $e_2$ and $e_3$ via a pitchfork bifurcation. The details of these mechanisms are further investigated in 
\ref{SD_to_PD_1}.}
\label{fig:SD2_1}
\end{figure*}
In Figure \ref{fig:SD2_1}, the behaviour of the dynamics of the Snowdrift game characterized by matrix~\eqref{matrix:snow2} is showcased.
For a small interval of values of delays, i.e. $\tau_C <0.00715429$, we can observe the coexistence of two internal stationary states that collide in a saddle-node bifurcation by increasing the delay magnitude $\tau_D$.
In the remainder of the space, the dynamics are similar to the ones observed for matrix~\eqref{matrix:snow1}. 


\subsection{Prisoner's Dilemma game}
Lastly, we analyse the Prisoner's Dilemma. 
In this game, each player can choose to incur a cost $c$ to provide the opponent with a benefit $b$ (C) or not (D) \cite{doebeli:el:2005}.
Additionally, here, we introduce a base endowment of $c$ so that all of the possible payoffs of the game are non-negative. Hence, the game is represented by the following matrix:
\begin{equation}\label{matrixgenPD}
\begin{array}{c}
\\
 C \\
 D 
\end{array}
\begin{array}{c}
\begin{matrix}
C & \! & \!& \! & \!D 
\end{matrix} \\
\left(\ \begin{matrix}
R=b & S=0 \\
T=b+c & P=c \\
\end{matrix}\ \right)
\end{array}
\end{equation}
where $b>c>0$. 
If no delays are present, the game has one stable stationary state, full defection ($x_0=0$), no internal stationary state exists, and full cooperation ($x_1=1$) is unstable.
For the Prisoner's Dilemma (\eqref{matrixgenPD}) the system~\eqref{compartment} becomes:
\begin{align}\label{compartmentpd}
\frac{dx(t)}{dt} =& \frac{y_C(t)(1-x(t))}{\tau_C}-\frac{y_D(t)x(t)}{\tau_D} \nonumber \\
\frac{dy_C(t)}{dt} =& y_C \left(\frac{\tau_C-1}{\tau_C}-\frac{y_C(t)}{\tau_C}-\frac{y_D (t)}{\tau_D}\right)+b x^2(t) \\
\frac{dy_D(t)}{dt} =& y_D\left(\frac{\tau_D-1}{\tau_D}-\frac{y_C(t)}{\tau_C}-\frac{y_D (t)}{\tau_D}\right)+ \nonumber\\
&(b x(t) +c) (1-x(t)) \nonumber
\end{align}
In the interval $[0,1]$, three stationary states of the system~\eqref{compartmentpd} are possible.
Full defection (equivalent to $x_0$) becomes:
\begin{equation}
e_0=\left(0, 0, \frac{1}{2} \left(\tau_D-1+\sqrt{4 c \tau_D+\tau_D^2-2 \tau_D+1}\right)\right).
\end{equation}
In $e_0$, the condition~\eqref{lambdastar} becomes
\begin{equation}
\frac{1}{2\tau_D} \left(\tau_D-1+\sqrt{4 c \tau_D+\tau_D^2-2 \tau_D+1}\right)>1.
\end{equation}
For the condition to hold, the inequality $c \geq 1$ needs to be true.
The full cooperation stationary state (equivalent to $x_1$) takes the following form:
\begin{equation}
e_1=\left(1, \frac{1}{2} \left(\tau_C-1+\sqrt{4 b \tau_C+\tau_C^2-2 \tau_C+1}\right), 0\right).
\end{equation}
Then, in $e_1$ the condition~\eqref{lambdastar} becomes 
\begin{equation}
\frac{1}{2\tau_C} \left(\tau_C-1+\sqrt{4 b \tau_C+\tau_C^2-2 \tau_C+1}\right)>1,
\end{equation}
which holds true when $b\geq 1$.
Additionally, an internal stationary state appears:
\begin{equation}
e_2 = \left(x_2^*, -\frac{\tau_C x_2^* y_D^*}{\tau_D (x_2^*-1)}, y_D^*\right) 
\end{equation}
where 
\begin{align}
y_D^* =& -\frac{1}{2} (x_2^*-1) \nonumber \\
&\left(\tau_D-1+\sqrt{\tau_D (4 b x_2^*+4 c-2)+\tau_D^2+1}\right) \\
x_2^*=& \frac{1}{2} \left(\frac{-2 c \tau_C+\tau_C-\tau_D}{b \tau_C-b \tau_D}+\sqrt{\frac{-4 c+\tau_C-\tau_D}{b^2 (\tau_C-\tau_D)}}\right).
\end{align}
To ensure that the population grows exponentially, we have 
\begin{equation}
 -\frac{\tau_C x_2^* y_D^*}{\tau_D\tau_C (x_2^*-1)}+\frac{y_D^*}{\tau_D}>1,
\end{equation}
and $x_2^*>1/b$.
Hence, for the population to grow in each stationary state, an additional constraint on the game has to be introduced: $c\geq1$. 
In the internal stationary state, the following needs to hold: $x_2^*>1/b$.

\textit{Homogeneous stationary states.}
The stability analysis of the full defection stationary state $e_0$ shows that it always exists and is always stable. 
Full cooperation $e_1$ is unstable, unless the following is true: $\tau_D > c/(-b+b^2) \land \tau_C < r$ where $r = \frac{2 (b-1) b \tau_D-c \left(-2 b \tau_D+\tau_D+1+\sqrt{\tau_D (4 b+4 c+\tau_D-2)+1}\right)}{2 (b+c-1) (b+c)}$. 
Full cooperation becomes stable for a big enough delay of defectors and a small enough delay of cooperators.
Details of the stability analysis are presented in \ref{SA_PD}.

\textit{Internal stationary state.}
The system does not have an internal stationary state if no delays are present.
Then, for $\tau_D > c/(-b+b^2)$ and $\tau_C = r$ the internal stationary state appears at  $x_2^* =1$.
A further increase in $\tau_D$ leads to a decrease in $x_2^*$.
An increase in $\tau_C$ always leads to an increase in $x_2^*$ until it disappears again when $\tau_C >r$.
Notably, $x_2^*$ attains positive values only if $\tau_D>\tau_C$.
In the limit of defector delay approaching infinity, $\tau_D \to \infty$, $x_2^*$ approaches a limiting value $1/b$, ensuring population growth in the internal stationary state.

\textit{One delay present.}
We analyse the system's behaviour when only one of the delays is present. For $\tau_C =0$ the system~\eqref{compartmenttc0} has one internal stationary state in the interval  $(0, 1)$, which coincides with the limit of the internal stationary state $x_2^*$:

$$\tilde{x}_2 = \frac{1+\sqrt{\frac{4 c}{\tau_D}+1}}{2 b} = \lim_{\tau_C \to 0} x_2^*.$$

In the case of no defector delay ($\tau_D = 0$) the system~\eqref{compartmenttd0} does not have an internal solution in the $(0, 1)$ interval. This result is explained by the fact that the existence of the internal stationary state of the general Prisoner's Dilemma depends on the presence of the defector delay. Consequently, we see that in the limit of $\tau_D \to 0$, the internal solution takes a value that is always less than 0: $\lim_{\tau_D \to 0} x_2^* = \frac{-2 c+1+b \sqrt{\frac{\tau_C-4 c}{b^2 \tau_C}}}{2 b}<0$.

We show that for the Prisoner's Dilemma in the cost-benefit form (transformed for all payoffs to be non-negative), complete defection is always stable, regardless of the delays. 
However, it is possible for full cooperation to become stable and for the unstable internal stationary state to appear. 
If both extreme stationary states are stable, an increase in the strategy's delay leads to a decrease in the size of the basin of attraction of the respective stationary state.

\textit{Example.}
For the numerical analysis, we use the following payoff matrix:

\begin{equation}\label{matrix:prisoner}
\begin{array}{c}
\\
 C \\
 D 
\end{array}
\begin{array}{c}
\begin{matrix}
C & \! & \!& \! & \!D 
\end{matrix} \\
\left(\ \begin{matrix}
R=3 & S=0 \\
T=5 & P=2 \\
\end{matrix}\ \right)
\end{array}
\end{equation}
\begin{figure*}
 \centering
\includegraphics[width = \textwidth]{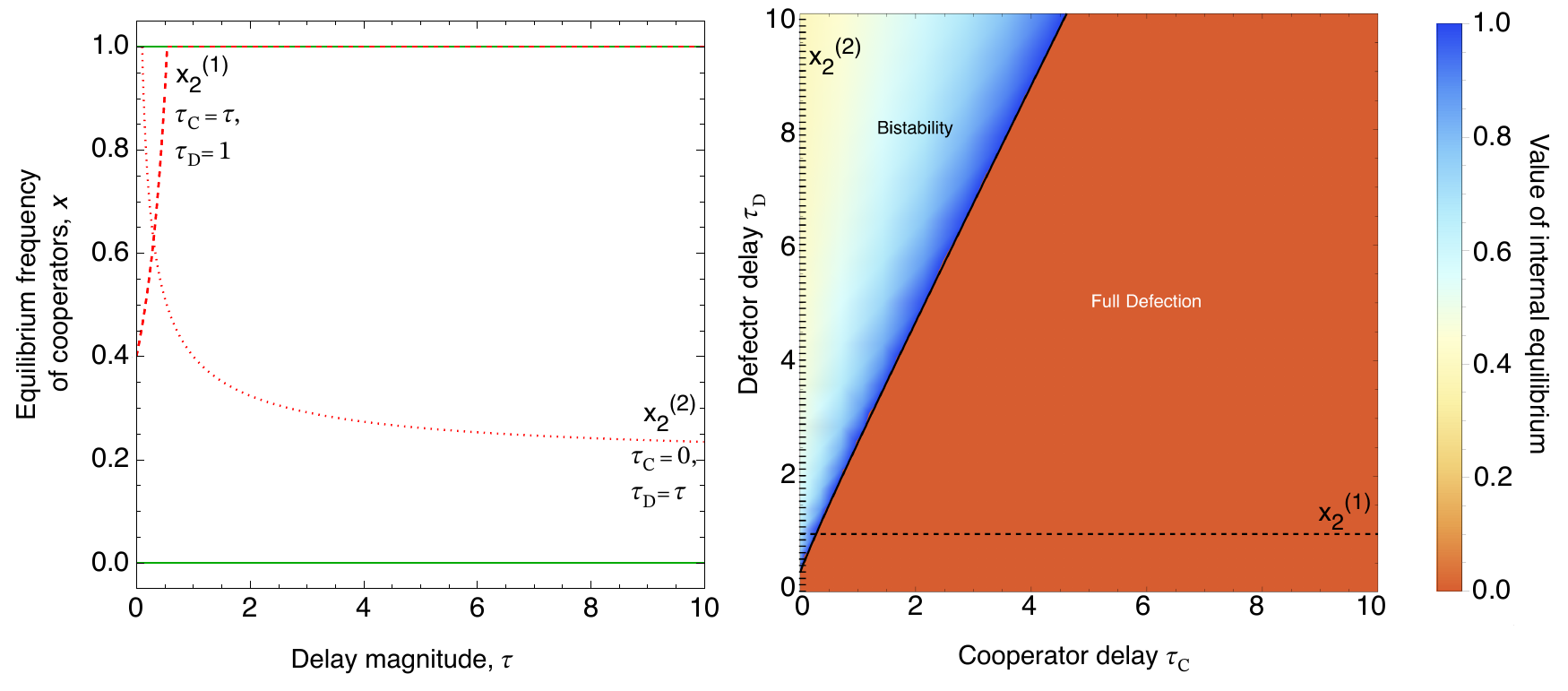}
\caption{
Stability of the stationary states of the Prisoner's Dilemma represented by matrix~\eqref{matrix:prisoner}. 
On the left, the stationary state values for specific values of delays are plotted.
The dashed line represents the internal stationary state $x_2^{(1)}$ as a function of $\tau$, when $\tau_C = \tau$, $\tau_D = 1$. An increase in $\tau$ leads to an increase in stationary state value. The internal stationary state ceases to exist for a high enough value of $\tau$, and full cooperation is not stable anymore.
The dotted line represents internal stationary state $x_2^{(2)}$ as a function of $\tau$, when $\tau_C = 0, \tau_D = \tau$. Full defection is the only stable stationary state for small values of $\tau$, and no internal stationary state exists. With the increase in $\tau$, the internal stationary state appears, and full cooperation becomes stable. The value of the internal stationary state decreases with the increase of $\tau$.
On the right, the stability of the system in the parameter space $\tau_C$ and $\tau_D$ is shown. 
The solid black line indicates the point of the bifurcation. In the "Full Defection" region, only $e_0$ is stable. In the "Bistability" region, both $e_0$ and $e_1$ are stable, and their basins of attraction are divided by the unstable internal stationary state. The colour indicates the value of the internal stationary state. The values considered on the left are represented by dashed ($e_2^{(1)}$) and dotted ($e_2^{(1)}$) lines, respectively.
The effects of only one delay present can be observed on the left and bottom edges of the plot.}
\label{fig:PD}
\end{figure*}
Figure \ref{fig:PD} explores the change in the system stability in the parameter space of delays. 
In the majority of the parameter space full defection is the only stable stationary state. 
As the delay of defectors increases, full cooperation becomes stable, as indicated by the colourful region in the plot, the colours representing the value of the internal stationary state.

\subsection{Game transitions}\label{gametransition}
\begin{figure*}
\centering
\includegraphics[width=0.8\columnwidth]{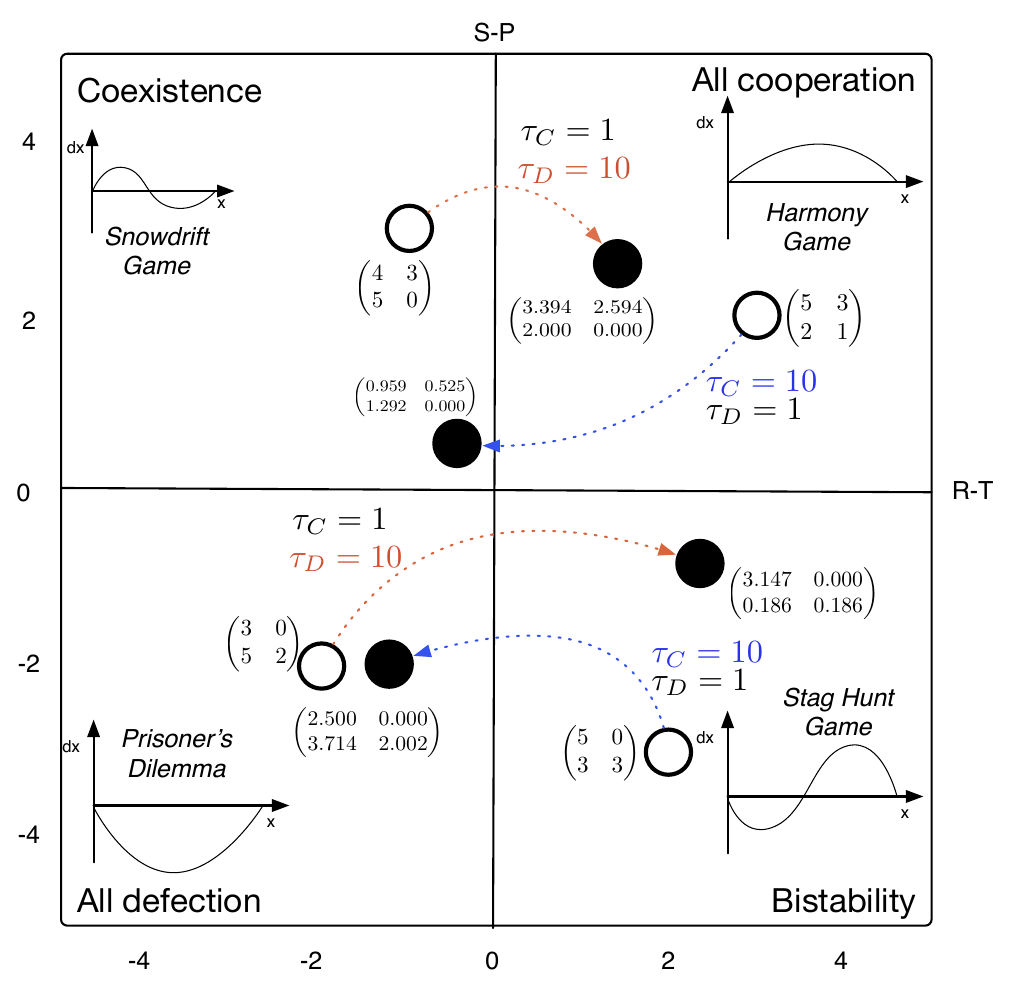}
\caption{The games described by the general Matrix~\eqref{matrix:general} can be classified into four classes depending on the sign of the advantage of cooperators playing against cooperators ($R-T$) and the advantage of defectors against defectors ($S-P$).
The introduction of strategy-dependent delays changes the effective game played in the population. With high enough values of delays present, the Stag-Hunt game becomes a Prisoner's Dilemma and vice-versa. An analogous transition can be observed between the Snowdrift and Harmony games.
The open circles represent the original games as discussed in Section \ref{results}. The filled circles show the effective game obtained by introducing delays, as indicated on the arrows. 
An increase in cooperator delay leads to a transition from right to left of the coordinate system and an increase in defector delay - from left to right.
The possible transitions between the games are limited by the assumptions made on the forms of the games.}\label{fig:discussion}
	\end{figure*}

In our analysis, we show that introducing delays may not only influence the game's properties but also change the type of the game entirely. Introducing strategy-specific delay is detrimental to the strategy affected. 
Thus, with a high enough cooperator delay, a Stag-Hunt may effectively become a Prisoner's Dilemma. And vice-versa, a defector delay may transform a Prisoner's Dilemma into a Stag-Hunt game. Similarly, the transition between a Snowdrift and Harmony (a game where all cooperation is always stable) can be achieved. 
The possible game transitions are represented in Figure~\ref{fig:discussion}, with open circles representing the original games and the closed ones showing the effective games played in the presence of delays. 
In order to recover the effective game payoff matrix for specific delay values, we need to obtain the numerical values of $\frac{dx(t)}{dt}$ given in~\eqref{compartment} depending on $x(t)$, assuming $\frac{dy_i(t)}{dt}=0$ for $i \in \{C, D\}$.
Then, the parameters of the effective game are determined by fitting the replicator equation $x(t)(1-x(t))(U_C(x)-U_D(t))$ to the obtained data, using the least-squares fit.

This result emphasizes the importance of delays in the analysis of games and the impact they have on the evolutionary dynamics.
Notably, game transitions presented in Figure~\ref{fig:discussion} result from cost-benefit forms of the games used in the results section~\ref{results}. 
However, other transitions can be observed by considering the general form of the payoff matrix~\eqref{matrix:general}.
In particular, a change between the Prisoner's Dilemma and the Snowdrift occurs, showing the possibility of a vertical shift in the coordinate system.
A numerical analysis of a possible transition is presented in~\ref{SD_to_PD_1}. 

\section{Conclusions}\label{conclusions}

Differential equations with time delays are infinite-dimensional dynamical systems, and therefore, any analytical solutions are difficult to obtain. 
Here, we presented a different approach to model time delays in replicator dynamics of evolutionary games. 
In our Kindergarten model, strategy-dependent time delays correspond to rates at which offspring leave the corresponding kindergarten and are able to play games. Our qualitative results coincide with conclusions presented in \cite{miekisz:PRE:2021} while providing benefits of lower complexity and more accessible analysis.

We derive explicit formulas for the stationary fraction of cooperators in the population.
Namely, we show that in the Stag-Hunt game, a strategy delay leads to a decrease in the size 
of its basin of attraction of the corresponding absorbing state. 
High enough cooperator time delay destabilises full cooperation. The same cannot be said about full defection, which is stable regardless of delay values.
Similar behaviour can be observed in the Prisoner's Dilemma game. 
With the introduction of time delays, full cooperation can become locally asymptotically stable.
Again, it is not possible to destabilise full defection. 


In both games, cooperation can be leveraged when we introduce stochasticity in finite populations. According to the ``one-third rule'' \cite{nowak:Nature:2004}, generalized in~\cite{nalecz:PRL:2018}, a strategy can be evolutionarily stable in finite populations if its basin of attraction is bigger than $1/3$ (the fixation probability of the competing strategy is smaller than the inverse of the population size). Therefore, by increasing the basin of attraction, delays can give the cooperators the needed advantage to take over the population.
The Snowdrift game represents a reverse situation 
to the Stag-Hunt and Prisoner's Dilemma ones - full defection can never be stable. 
However, full cooperation can become stable. If the internal stationary state exists, it is stable, and its value depends on the delays. 
An increase in the delay of a given strategy decreases its fraction in the coexistence stationary state. 
Moreover, an additional second internal stationary state may appear.

The effects of delays are not limited to shifting internal stationary states within the game class, but additionally, they can lead to a change in the nature of the game itself. 
We show that introducing strategy-dependent delays alters the effective games played and subsequently leads to a change of optimal strategies. 
This result underlines the importance of considering the temporal structure of studied systems.
Our approach can be used in games with multiple strategies and multi-player games.
 
\section*{Competing interests}
The author(s) declare no competing interests.

\section*{Acknowledgements}

MF thanks Javad Mohamadichamgavi, Péter Bayer, and  Stefano Giaimo for the fruitful discussions.
Funding from the Max Planck Society is graciously acknowledged. This research is supported by the European Union’s Horizon 2020 research and innovation programme under the Marie Skłodowska-Curie grant under the grant agreement number 955708. 

\section*{Code availability}
Appropriate computer code describing the model and used to produce the figures is available on Github at
\url{https://github.com/tecoevo/compartmentalised_timedelays}.

\appendix

\section{Kindergarten model derivation}\label{modelderivation}

In the Kindergarten model, a large but finite population is considered. 
Individuals in the population take part in pairwise interactions, play games, and obtain payoff. 
The payoff is, in turn, translated into several identical offspring created. 
The newly created offspring are placed in a strategy-specific kindergarten compartment and rejoin the adult population upon maturation, with per capita rate $1/\tau_i$ for $i \in \{C, D\}$.

We assume that in a small time interval $\epsilon$ only a fraction $\delta_K(\epsilon)$ of juveniles in the kindergarten can mature, with $\delta_K$ being a function such that $\frac{d\delta_K(\epsilon)}{d\epsilon} \geq 0$, $\delta_K(0)=0$, $\frac{d\delta_K(0)}{d\epsilon}=const$. 
That is, in a given time interval, at least as many individuals have to mature as in the shorter interval; no juveniles can mature instantly.
Similarly, in a time interval $\epsilon$ only a $\delta_P(\epsilon)$ fraction of the adult population takes part in the game with $\delta_P$ being a function such that $\frac{d\delta_P(\epsilon)}{d\epsilon} \geq 0$, $\delta_P(0)=0$ and $\frac{d\delta_K(0)}{d\epsilon}=const$. 
That is, any interactions that have taken place in a shorter time interval have also occurred in a longer time, and no interactions take place within no time.
Then, the size of the kindergarten of strategy $i$ at time $t+\epsilon$ is equal to:
	\begin{equation}
		k_i (\hat{t}+\epsilon) = \left(1-\frac{\delta_K(\epsilon)}{\tau_i}\right)k_i(\hat{t}) + \delta_P(\epsilon) p_i (\hat{t}) U_i (\hat{t}).
	\end{equation}
Similarly, the change in the size of the adult population using strategy $i$ can be calculated as:
	\begin{equation}
		p_i(\hat{t}+\epsilon) = (1-\delta_P(\epsilon))p_i(\hat{t}) + \frac{\delta_K(\epsilon)}{\tau_i} k_i (\hat{t}).
	\end{equation}
The size of the whole adult population follows:
	\begin{equation}
		p(\hat{t}+\epsilon) = (1-\delta_P(\epsilon))p(\hat{t}) + \frac{\delta_K(\epsilon)}{\tau_C} k_C (\hat{t})+\frac{\delta_K(\epsilon)}{\tau_D} k_D (\hat{t}).
	\end{equation}

For ease of calculations, we introduce variables $y_i (\hat{t})= k_i (\hat{t})/p(\hat{t}) $ for $ i \in \{C, D\}$, denoting the size of each kindergarten compartment relative to the whole adult population. Notably, the sum of $y_i (\hat{t})$ for all $i$ does not have to be equal to 1, since the variables are not fractions of the population.
Now we can derive equation for change in $x(\hat{t})$, $y_C(\hat{t})$ and $y_D (\hat{t})$.
\begin{align}
		x(\hat{t}+\epsilon) - x(\hat{t}) = & \frac{(1-\delta_P(\epsilon))p_C(\hat{t})+\frac{\delta_K(\epsilon)}{\tau_C}k_C(\hat{t})}{(1-\delta_P(\epsilon))p(\hat{t}) +\frac{\delta_K(\epsilon)}{\tau_C}k_C(\hat{t})  +\frac{\delta_K(\epsilon)}{\tau_D}k_D(\hat{t})} - \frac{p_C(\hat{t})}{p(\hat{t})} \\
	y_i(\hat{t}+\epsilon) - y_i (\hat{t}) = & \frac{(1-\frac{\delta_K(\epsilon)}{\tau_i})k_i(\hat{t}) + \delta_P(\epsilon) p_i (\hat{t}) U_i (\hat{t})}{(1-\delta_P(\epsilon))p(\hat{t}) +\frac{\delta_K(\epsilon)}{\tau_C}k_C(\hat{t})  +\frac{\delta_K(\epsilon)}{\tau_D}k_D(\hat{t})} - \frac{k_i(\hat{t})}{p(\hat{t})}
\end{align}

After some rearrangements, we get:

\begin{align}
x(\hat{t}+\epsilon) - x(\hat{t}) =& \delta_K(\epsilon)\frac{\frac{1}{\tau_C}k_C(\hat{t})(p(\hat{t})-p_C(\hat{t}))  -p_C(\hat{t})\frac{1}{\tau_D}k_D(\hat{t})}{p(\hat{t})\left((1-\delta_P(\epsilon))p(\hat{t}) +\frac{\delta_K(\epsilon)}{\tau_C}k_C(\hat{t})  +\frac{\delta_K(\epsilon)}{\tau_D}k_D(\hat{t})\right)} \\
y_i(\hat{t}+\epsilon) - y_i ( \hat{t}) =& \delta_P\frac{(\epsilon)p(\hat{t})\left(k_i(\hat{t})+p_i(\hat{t})U_i(\hat{t})\right)}{(1-\delta_P(\epsilon))p^2(\hat{t}) +\delta_K(\epsilon)p(\hat{t})\left(\frac{k_C(\hat{t})}{\tau_C}+\frac{k_D(\hat{t})}{\tau_D}\right)} \\
&\ - \delta_K(\epsilon)\frac{k_i(\hat{t})(\frac{p(\hat{t})}{\tau_i}+\frac{k_C(\hat{t})}{\tau_C}+\frac{k_D(\hat{t})}{\tau_D})}{(1-\delta_P(\epsilon))p^2(\hat{t}) +\delta_K(\epsilon)p(\hat{t})\left(\frac{k_C(\hat{t})}{\tau_C}+\frac{k_D(\hat{t})}{\tau_D}\right)}. \nonumber
\end{align}
Then, we divide both sides of the equations by $\epsilon$: 
 	\begin{align}\label{derivexy}
	\frac{x(\hat{t}+\epsilon) - x(\hat{t})}{\epsilon} =& \frac{\delta_K(\epsilon)}{\epsilon}\frac{\frac{1}{\tau_C}k_C(\hat{t})(p(\hat{t})-p_C(\hat{t}))-p_C(\hat{t})\frac{1}{\tau_D}k_D(\hat{t})}{p(\hat{t})\left((1-\delta_P(\epsilon))p(\hat{t}) +\frac{\delta_K(\epsilon)}{\tau_C}k_C(\hat{t})  +\frac{\delta_K(\epsilon)}{\tau_D}k_D(\hat{t})\right)} \\
	\frac{y_i(\hat{t}+\epsilon) - y_i (\hat{t})}{\epsilon} = & \frac{\delta_P(\epsilon)}{\epsilon}\frac{p(\hat{t})(k_i(\hat{t})+p_i(\hat{t})U_i(\hat{t}))}{(1-\delta_P(\epsilon))p^2(\hat{t}) +\delta_K(\epsilon)p(\hat{t})\left(\frac{k_C(\hat{t})}{\tau_C}+\frac{k_D(\hat{t})}{\tau_D}\right)}  \\
& - \frac{\delta_K(\epsilon)}{\epsilon}\frac{k_i(\hat{t})(\frac{p(\hat{t})}{\tau_i}+\frac{k_C(\hat{t})}{\tau_C}+\frac{k_D(\hat{t})}{\tau_D})}{(1-\delta_P(\epsilon))p^2(\hat{t}) +\delta_K(\epsilon)p(\hat{t})\left(\frac{k_C(\hat{t})}{\tau_C}+\frac{k_D(\hat{t})}{\tau_D}\right)} \nonumber
\end{align}
We recall that $\delta_K(0) = 0$ and $\delta_P(0) = 0$, hence, we can rewrite the above as:
\begin{align}
	\frac{x(\hat{t}+\epsilon) - x(\hat{t})}{\epsilon} = & \frac{\delta_K(0+\epsilon)-\delta_K(0)}{\epsilon} \frac{\frac{1}{\tau_C}k_C(\hat{t})(p(\hat{t})-p_C(\hat{t}))-p_C(\hat{t})\frac{1}{\tau_D}k_D(\hat{t})}{p(\hat{t})\left((1-\delta_P(\epsilon))p(\hat{t}) +\frac{\delta_K(\epsilon)}{\tau_C}k_C(\hat{t})  +\frac{\delta_K(\epsilon)}{\tau_D}k_D(\hat{t})\right)}
	\end{align}

\begin{align}
		\frac{y_i(\hat{t}+\epsilon) - y_i (\hat{t})}{\epsilon} = & \frac{\delta_P(0+\epsilon)-\delta_P(0)}{\epsilon}  \frac{p(\hat{t})(k_i(t)+p_i(\hat{t})U_i(\hat{t}))}{(1-\delta_P(\epsilon))p^2(\hat{t}) +\delta_K(\epsilon)p(\hat{t})\left(\frac{k_C(\hat{t})}{\tau_C}+\frac{k_D(\hat{t})}{\tau_D}\right)} \\
& - \frac{\delta_K(0+\epsilon)-\delta_K(0)}{\epsilon}\nonumber \frac{k_i(\hat{t})(\frac{p(\hat{t})}{\tau_i}+\frac{k_C(\hat{t})}{\tau_C}+\frac{k_D(\hat{t})}{\tau_D})}{(1-\delta_P(\epsilon))p^2(\hat{t}) +\delta_K(\epsilon)p(\hat{t})\left(\frac{k_C(\hat{t})}{\tau_C}+\frac{k_D(\hat{t})}{\tau_D}\right)} \nonumber
	\end{align}

Then, we perform the limit $\epsilon \to 0$ and obtain the following differential equations:

 	\begin{align}
		\frac{dx(\hat{t})}{d\hat{t}} =& \frac{d\delta_K(0)}{d\epsilon}\frac{\frac{1}{\tau_C}k_C(\hat{t})(p(\hat{t})-p_C(\hat{t}))-p_C(\hat{t})\frac{1}{\tau_D}k_D(\hat{t})}{p^2(\hat{t})} \\	\frac{dy_i(\hat{t})}{d\hat{t}} = & \frac{d\delta_P(0)}{d\epsilon}\frac{p(\hat{t})(k_i(\hat{t})+p_i(\hat{t})U_i(\hat{t}))}{p^2(\hat{t})} \\
& -\frac{d\delta_K(0)}{d\epsilon}\frac{k_i(\hat{t})(\frac{p(\hat{t})}{\tau_i}+\frac{k_C(\hat{t})}{\tau_C}+\frac{k_D(\hat{t})}{\tau_D})}{p^2(\hat{t})} \nonumber
	\end{align}

Now we denote $\frac{d\delta_P(0)}{d\epsilon}=Q$ and $\frac{d\delta_K(0)}{d\epsilon}=K$ and recall that $Q \geq 0$ and $K\geq 0$. 
Then, after rearrangements, we get

\begin{align}
		\frac{dx(\hat{t})}{d\textbf{}} = & K\left(\frac{y_C(\hat{t})(1-x(\hat{t}))}{\tau_C}  -\frac{x(\hat{t})y_D(\hat{t})}{\tau_D}\right) \nonumber \\
		\frac{dy_C(\hat{t})}{d\hat{t}} = &Q \left(y_C(\hat{t})+x(\hat{t})U_C(\hat{t})\right) \nonumber \\
& - K y_C(\hat{t})\left(\frac{1}{\tau_C}+\frac{y_C(\hat{t})}{\tau_C}+\frac{y_D(\hat{t})}{\tau_D}\right) \\
		\frac{dy_D(\hat{t})}{d\hat{t}} = & Q\left(y_D(\hat{t})+(1-x(\hat{t}))U_D(\hat{t})\right) \nonumber \\
& - K y_D\left(\hat{t})(\frac{1}{\tau_D}+\frac{y_C(\hat{t})}{\tau_C}+\frac{y_D(\hat{t})}{\tau_D}\right) \nonumber
\end{align}
For simplicity, in this work, we assume that $Q =K$.

Rescaling time by introducing
\begin{align*}
	t = a\hat{t},&\rightarrow \hat{t} = \frac{t}{a}.
\end{align*}
By the chain rule, we have
\begin{align*}
\frac{dx}{dt} &=\frac{dx}{d\hat{t}}\frac{d\hat{t}}{dt}=\frac{1}{a}\frac{dx}{d\hat{t}},\\
\frac{dy_C}{dt} &=\frac{dy_C}{d\hat{t}}\frac{d\hat{t}}{dt}=\frac{1}{a}\frac{dy_C}{d\hat{t}},\\
\frac{dy_D}{dt} &=\frac{dy_D}{d\hat{t}}\frac{d\hat{t}}{dt}=\frac{1}{a}\frac{dy_D}{d\hat{t}}.\\
\end{align*}
Setting $a = Q$ we obtain our final kindergarten model as in ~\eqref{compartment}.

 \begin{table*}
 \scriptsize{
 \centering
 \caption{Quantities and units}
 \begin{tabular}{c|c|c}
 quantity & unit & discription\\ \hline\hline
 $p_D$ & person & persons playing defect\\
 $p_C$ & person & persons playing cooperate\\
 $x$ & dim less & ratio of persons playing cooperate in $p$\\ \hline
 $k_D$ & person & kindergarden of playing defect\\
 $k_C$ & person & kindergarden of playing cooperate\\
 $y_D$ & dim less & relative size of the kindergarden $k_D$ with respect to the total population $p$\\
 $y_C$ & dim less & relative size of the kindergarden $k_C$ with respect to the total population $p$\\ \hline
 $1/\tau_D$ & dim less & defector maturation rate\\
 $\bar{\tau}_C$ & time & average cooperator maturation time \\
 $\bar{\tau}_D$ & time & average defector maturation time \\
 $1/\tau_C$ & dim less & cooperator maturation rate\\\
 $\epsilon$ & time& small time interval \\
 $\delta_P$ & dim less& Proportion of the adult population participating in the game in the time interval $\epsilon$ \\
 $\delta_K$ & dim less& fraction of kindergarten maturing in time interval $\epsilon$ \\
 $Q$ & 1/time & $\frac{d\delta_P(0)}{dt}$ \\
 $K$ & 1/time & $\frac{d\delta_K(0)}{dt}$ \\\hline
 $U_C$ & dim less & expected payoff of cooperator\\
 $U_D$ & dim  less & expected payoff of defector\\
 $R$, $S$, $T$, $P$ & dim less & payoffs defined by the payoffmatrix~\eqref{matrix:general}
 \end{tabular}
 }
 \label{tab:my_label}
 \end{table*}

\section{Stability Analysis}
\subsection{Stag-Hunt}\label{SA_SH}

We perform a stability analysis of the Stag-Hunt game's full defection stationary state $e_0$. Its eigenvalues are given by:

\begin{align}
	\lambda_{0,1} = & -\frac{\sqrt{4 b \tau_D+\tau_D^2-2 \tau_D+1}}{\tau_D}\\
\lambda_{0,2} = & -\frac{\tau_D-1+\sqrt{4 b \tau_D+\tau_D^2-2 \tau_D+1}}{2 \tau_D}\\
\lambda_{0,3} = & \frac{\tau_D+1- \sqrt{4 b \tau_D+\tau_D^2-2 \tau_D+1}}{2  \tau_D}-\frac{1}{ \tau_C}.\\
	\end{align}

This stationary state is always stable.

In contrast, the full cooperation stationary state $e_1$ can change its stability. The eigenvalues of the stationary state are given by:

\begin{align}
	\lambda_{1,1} = & -\frac{\sqrt{(4 a-2) \tau_C+\tau_C^2+1}}{\tau_C}\\
\lambda_{1,2} = & \frac{1- \sqrt{4 a \tau_C+\tau_C^2-2 \tau_C+1}}{2 \tau_C} \nonumber \\
&-\frac{1+\sqrt{4 b \tau_D+\tau_D^2-2 \tau_D+1}}{2 \tau_D}\\
\lambda_{1,3} = &\frac{1 -\sqrt{(4 a-2) \tau_C+\tau_C^2+1}}{2 \tau_C } \nonumber \\
&-\frac{1- \sqrt{4 b \tau_D+\tau_D^2-2 \tau_D+1}}{2  \tau_D}.
	\end{align}

The change in the stability of the stationary state happens via transcritical bifurcation when $\lambda_{1, 3} = 0$, which takes place when $\tau_C = m$.
At this point, we observe the internal stationary state reaching the full cooperation stationary state ($e_1 = e_2$) and the stability of the two stationary states exchanges. 
For $\tau_C\geq m$, the full cooperation stationary state $e_1$ is unstable, and $e_2$ does not exist. 
 When $e_2$ exists, it is unstable, and the full cooperation stationary state is stable.

 \subsection{Snowdrift Game}\label{SA_SG}

First, we present the stability analysis of the full cooperation stationary state $e_1$of the Snowdrift game. The eigenvalues are given by:

\begin{align}
	\lambda_{1,1} = &-\frac{\sqrt{4 b \tau_C-2 (c+1) \tau_C+\tau_C^2+1}}{\tau_C}\\
\lambda_{1,2} = & \frac{1 -\sqrt{4 b \tau_C-2 (c+1) \tau_C+\tau_C^2+1}}{2 \tau_C } \nonumber \\
& -\frac{1+\sqrt{(4 b-2) \tau_D+\tau_D^2+1}}{2  \tau_D}\\
\lambda_{1,3} = &\frac{ 1-\sqrt{4 b \tau_C-2 (c+1) \tau_C+\tau_C^2+1}}{2 \tau_C } \nonumber \\
&-\frac{1- \sqrt{(4 b-2) \tau_D+\tau_D^2+1}}{2  \tau_D}.
	\end{align}

 The change in stability of the stationary state happens via a transcritical bifurcation when $\lambda_{1, 3}=0$. The bifurcation takes place when $\tau_D=n$.
 At this point, we observe one of the internal stationary states ($e_2$ or $e_3$ depending on the parameter values) and the full cooperation stationary state exchange stability.
 For $\tau_D>n$, full cooperation is a stable stationary state.

 We perform the stability analysis of the full defection stationary state $e_0$. The eigenvalues are given by:

\begin{align}
	\lambda_{0,1} = & -\frac{| \tau_D-1| }{\tau_D}\\
\lambda_{0,2} = & -\frac{1+ \sqrt{\tau_C (4 b-4 c-2)+\tau_C^2+1}}{2 \tau_C } \nonumber \\
& +\frac{1-| \tau_D-1| }{2 \tau_D}\\
\lambda_{0,3} = &- \frac{1- \sqrt{\tau_C (4 b-4 c-2)+\tau_C^2+1}}{2 \tau_C } \nonumber \\
& +\frac{ 1-| \tau_D-1| }{2  \tau_D}.
	\end{align}

The stationary state changes its stability via a transcritical bifurcation when $\lambda_{0,3} = 0$, which occurs when $\left( b<c+1\land \tau_D=p \right)$. The full defection stationary state is stable if $\left( b<c+1\land \tau_D>p \right)$ and unstable otherwise. For the model assumptions not to be violated, we assume that $b<c+1$ and hence, the full defection stationary state $e_0$ is always unstable.
	
 \subsection{Prisoner's dilemma}\label{SA_PD}

We perform the stability analysis of the full defection stationary state $e_0$. The eigenvalues of the stationary state are given by:

\begin{align}
	\lambda_{0,1} = & -\frac{\sqrt{(4 c-2) \tau_D+\tau_D^2+1}}{\tau_D}\\
\lambda_{0,2} = & -\frac{\tau_D-1+\sqrt{(4 c-2) \tau_D+\tau_D^2+1}}{2 \tau_D}\\
\lambda_{0,3} = & \frac{1- \sqrt{4 c \tau_D+\tau_D^2-2 \tau_D+1}}{2 \tau_D}+\frac{\tau_C -2}{2 \tau_C}.
	\end{align}

The full defection stationary state is always stable.

Now, we perform the analysis of the full cooperation stationary state $e_1$. The eigenvalues are given by:

\begin{align}
	\lambda_{1,1} = & -\frac{\sqrt{(4 b-2) \tau_C+\tau_C^2+1}}{\tau_C}\\
\lambda_{1,2} = & -\frac{1+ \sqrt{\tau_D (4 b+4 c-2)+\tau_D^2+1}}{2 \tau_D} \nonumber \\
&  +\frac{1-\sqrt{(4 b-2) \tau_C+\tau_C^2+1}}{2 \tau_C}\\
\lambda_{1,3} = &-\frac{1- \sqrt{\tau_D (4 b+4 c-2)+\tau_D^2+1}}{2  \tau_D} \nonumber \\
&+\frac{1-\sqrt{(4 b-2) \tau_C+\tau_C^2+1}}{2 \tau_C }.
	\end{align}

Full cooperation can change its stability via a transcritical bifurcation when $\lambda_{1, 3} = 0$, which takes place when $\{\tau_D > c/(-b+b^2) \land \tau_C = r\}$. 
At the bifurcation point, we have $e_1=e_2$ and the internal stationary state exchanges stability with full cooperation stationary state. 
For $\tau_D > c/(-b+b^2) \land \tau_C < r$ full cooperation is stable. 
In this parameter region, the internal stationary state $e_2$ exists in the relevant interval ($x_2^* \in (1,0)$) and is unstable. 
Outside this parameter space, stationary state $e_2$ does not exist, and full cooperation is unstable.

\section{Transition from a Snowdrift to Prisoner's Dilemma}\label{SD_to_PD_1}

In Sec.~\ref{gametransition}, game transitions caused by time delays are discussed.
Here, we are interested in exploring a less restricted parameter regime in the form of a general payoff matrix \eqref{matrix:general} and observed game transitions.
In particular, we investigate the transition from the Snowdrift game into a Prisoner's Dilemma, meaning from a game (Snowdrift) that can exhibit two internal states~\cite{miekisz:PRE:2021} to the game of full defection, which can, in turn, become a bistability game.
This implies we are interested in the case when the inequality $P<S$ (Snowdrift) of the general payoff matrix becomes $P>S$ (Prisoner's Dilemma).
We use the payoff matrix~\eqref{matrix:snow2} as a case study, meaning if $P>1.1$, the Snowdrift game has become a Prisoner's Dilemma.

\begin{figure*}
\includegraphics[width = 0.9\textwidth]{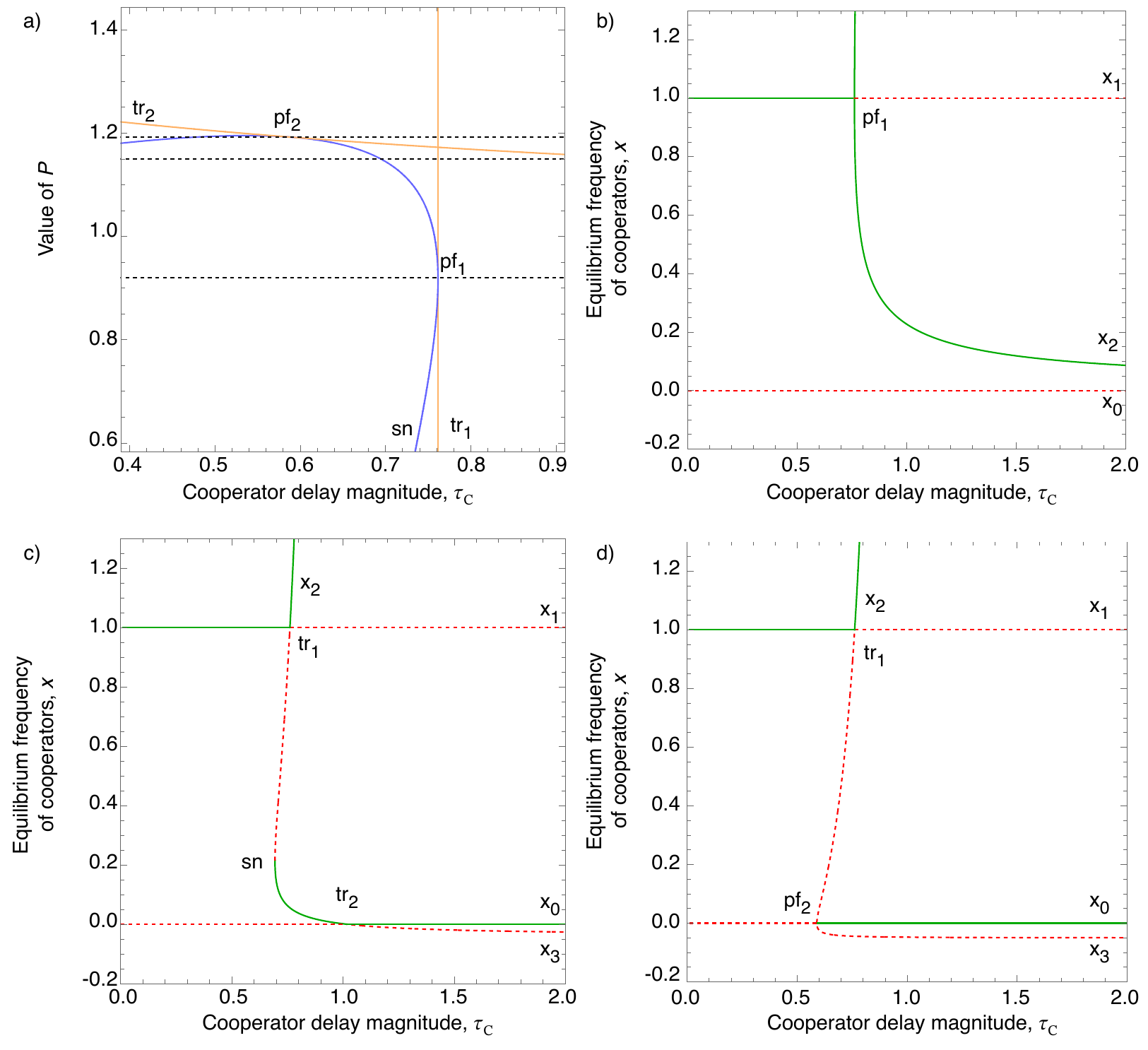}
 \caption{
 \textbf{a)} The saddle-node bifurcation $sn$ collides with the transcritical bifurcation $tr_1$ in a codimension two pitchfork bifurcation $pf_1$ and with the transcritical $tr_2$ in a supercritical pitchfork $pf_2$.
 \textbf{b)} At $P\approx 0.91624583$, $\tau_C\approx 0.761597$, the saddle-node bifurcation $sn$ enters the biologically relevant interval $[0,1]$ via a supercritical bifurcation pitchfork $pf_1$ giving rise to two internal mixed stationary states see \textbf{c)} and leaves the interval of interest via a supercritical pitchfork bifurcation at $P\approx1.19206938$ and $\tau_C=0.589708546$ (see \textbf{d)} ).
 The remaining parameter are $R=3.05$, $S=1.1$, $T=5$, $P=0$, and $\tau_D=2.0$.
 }
\label{fig:var_p}
\end{figure*}

We start by continuing the branch of mixed stationary states $e_2$ in Fig.~\ref{fig:SD1} outside of the interval $[0,1]$ i.e. $x_2>1$ and observe that at $\tau_C\approx0.65833239$ a saddle-node bifurcation $sn$ occurs, which gives rise to $e_2$ as depicted in Fig.~\ref{fig:SD1}.
Throughout the following discussion, we are particularly interested in the location of $sn$ and the associated changes in the effective game with time delays. We explore the $P,\tau_C$ parameter plane where $sn$ is depicted by the blue curve in Fig.\ref{fig:var_p} a).

At $P\approx 0.91624583$, and $\tau_C\approx 0.761597$ the fold bifurcation $sn$ enters the relevant domain $(0\leq x \leq 1)$ by colliding with the transcritical bifurcation $tr_1$ forming a codimension 2 supercritical pitchfork bifurcation $pf_1$; see Fig.\ref{fig:var_p} b).
Further increasing of $P$ results in the unfolding of the pitchfork bifurcation $pf_1$ and the formation of two internal stationary states like Fig.~\ref{fig:SD2_1}.
It appears that such configuration is quite robust in the parameter space as already indicated in Fig.~\ref{fig:SG_params} and noted by Mi\c{e}kisz and Bodnar~\cite{miekisz:PRE:2021} in the original DDE model.
By further increasing $P$, we observe a second transcritical bifurcation $tr_2$ for sufficiently large time delay $\tau_C$ as indicated by the corresponding one-dimensional bifurcation diagram for $P=1.15$ in Fig.\ref{fig:var_p} c).
Notably, we are already in the Prisoner's Dilemma parameter domain, but due to the time delay $\tau_C$, we still have two internal stationary states similar to the Snowdrift game.
As indicated by Fig.\ref{fig:var_p} a), this behaviour is robust until the saddle-node leaves the interesting interval $[0, 1]$.
If $P$ is further increased, $sn$ leaves the interesting interval $[0,1]$, similar to the pitchfork $pf_1$ caused by the collision of $sn$ and $tr_1$.
Indeed, $sn$ collides for $P\approx1.19206938$ and $\tau_C\approx0.589708546$ with the previously observed $tr_2$ at a subcritical pitchfork bifurcation at $\tau_D\approx0.589708546$ (see \ref{fig:var_p} d) ). 

 \section{Extinction in a Snowdrift game due to time delay}\label{Ex_SD}

\begin{figure*}
\centering
\includegraphics[width = 0.9\textwidth]{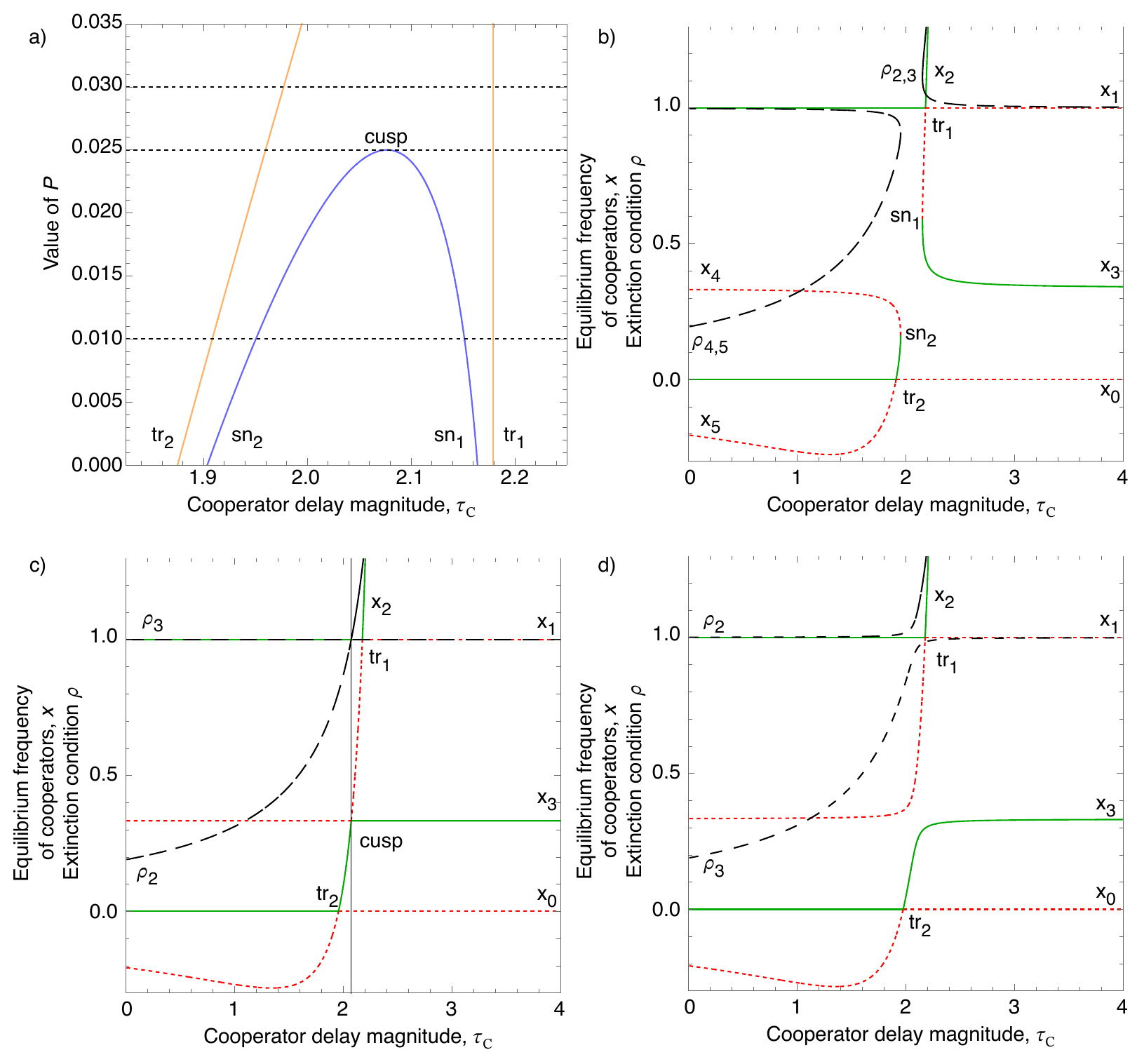}
 \caption{
\textbf{a)} The saddle-node bifurcations $sn_1$ and $sn_2$ colide at a cusp bifurcation.
\textbf{b)} The branches $e_4$ and $e_5$ are associated with an decreasing population since $\rho_{4,5}(\tau_C)<1$ (black dashed line) while the branches $e_2$ and $e_3$ are associated with an increasing population since $\rho_{2,3}(\tau_C)>1$ (black dashed line).
\textbf{c)} At the cusp bifurcation, the population of $e_3$ remains constant since $\rho_3 (\tau_C)=1$, and the growth of population of $e_2$ depends on the time delay $\tau_C$, i.e. for $\tau_C<2.07708$ the population goes extinct and grows otherwise.
\textbf{d)} The detaching of $e_2$ and $e_3$ leads to an extinct population or sufficiently large enough time delay, since for $\tau_C>2.17929238$ only $e_3$ remains as a single attractor in the interval $[0,1]$ and is associated with a decreasing population.
 The remaining parameter are $R=2.0$, $S=0.5$, $T=2.95$, and $\tau_D = 5.0$.
 }
\label{fig:cusp_p}
\end{figure*}

In \ref{SD_to_PD_1}, we investigated the transition from Snowdrift game to Prisoner's Dilemma and the change in the associated bifurcation structure, which depended on the location of the saddle-node bifurcations $sn$.
Here, we consider a similar scenario of investigating the dynamics of changing $P$ in the following general payoff matrix
\begin{equation}\label{matrix:cusp}
\begin{array}{c}
\\
 C \\
 D 
\end{array}
\begin{array}{c}
\begin{matrix}
C & \! & \!& \! & \!D 
\end{matrix} \\
\left(\ \begin{matrix}
R=2.0 & S=0.5 \\
T=2.95 & P=0.01 \\
\end{matrix}\ \right).
\end{array}
\end{equation}
In the following analysis, we always vary $\tau_C$ and keep $\tau_D=5.0$ fixed. 
The corresponding bifurcation diagram, in Fig.~\ref{fig:cusp_p} b), depicts the x-component of the respective branch of stationary states.
First, we note that both trivial stationary states $e_0$ and $e_1$ exchange stability at transcritical bifurcation $tr_1$ and $tr_2$ respectively.
In particular, the scenario at $tr_1$ is similar to Fig.~\ref{fig:SD2_1}, as the associated branch $e_2$ will collide at the saddle-node $sn_1$ with $e_3$ giving rise to two internal stationary states.
The extinction condition Eq.~\eqref{lambdastar} $\rho_{2,3}(\tau_C,\tau_D=5.0)$ (see the corresponding black dashed line) of both internal branches $e_2$, $e_3$ is larger than 1, implying the growth of underlying population.

However, we also observe an additional branch $e_5$ associated with $tr_2$.
This branch bifurcates with $e_4$ at a second saddle-node $sn_2$.
As indicated by $\rho_{4,5}(\tau_C,\tau_D=5.0)$, the population of both branches goes extinct since $\rho_{4,5}<1$.
Therefore, by increasing the time delay $\tau_C$, we note the effective game for Matrix~\eqref{matrix:cusp} transition from bistability (separated by the repelling branch $x_4$) to a Prisoner's Dilemma to coexistence.

Similar to \ref{SD_to_PD_1}, we increase $P$ of the payoff matrix~\eqref{matrix:cusp}, which results in the collision of $sn_1$ and $sn_2$ at a cusp bifurcation $\tau_C \approx2.07708$, $P=0.025$ as depicted in the two-dimensional bifurcation diagram in Fig.~\ref{fig:cusp_p} a).
At the cusp bifurcation for $P=0.025$ (see Fig.~\ref{fig:cusp_p} c), $sn_1$ and $sn_2$ are connected via a transcritical bifurcation for varying $\tau_C$.
For this degenerated case, the population of the connecting branch $x_3$ remains constant since $\rho_{3}=1$.  
Additionally, the population of $x_2$ decreases if $\tau_C < 2.07708$ since $\rho_{2}<1$ and increases for $ 2.07708<\tau_C< 2.17929239$ until $x_2$ leaves the relevant interval at $tr_1$ with $\tau_C \approx 2.17929239 $.

Further increase of $P$ results in a detaching of $x_2$ and $x_3$.
The $P = 0.03$ case is depicted in Fig.~\ref{fig:cusp_p} d).
The branch $x_2$ is repelling separating cooperating and defection for $\tau_C<1.97722372$, denoting $tr_2$ and $x_3$ entering the $[0, 1]$ interval.
However, $\rho_{3}(\tau_C)<1$ and the stationary state $e_3$ is attractive, meaning the associated population goes extinct.
In other words, in this scenario, an increase in the time delay causes the population to go extinct. 

\clearpage
 \bibliographystyle{elsarticle-num} 






\end{document}